\documentclass[a4paper,11pt]{article}
\usepackage{cite}
\usepackage[all]{xy}
\usepackage{amsmath,amsfonts,amssymb,amsthm,epsfig,amscd,comment,latexsym,psfrag}
\usepackage{CJK}
\usepackage{mathrsfs}
\usepackage{nicefrac,xspace,tikz}
\usepackage{arydshln}
\usepackage{stmaryrd}
\usepackage{graphicx,bm}
\usepackage{amscd}
\usetikzlibrary{arrows}

\makeatletter

\newcommand{\Rmnum}[1]{\expandafter\@slowromancap\romannumeral #1@}
\makeatother
\usepackage{graphicx,amssymb,amsmath,bm,latexsym}
\allowdisplaybreaks

\begin{document}
	
\begin{titlepage}
\begin{center}
{\Large\bf Generalized $\beta$ and $(q,t)$-deformed partition functions with $W$-representations
and Nekrasov partition functions}\vskip .2in
{\large
Fan Liu$^{a,}$\footnote{liufan-math@cnu.edu.cn},
Rui Wang$^{b,}$\footnote{wangrui@cumtb.edu.cn},
Jie Yang$^{a,}$\footnote{yangjie@cnu.edu.cn}
and
Wei-Zhong Zhao$^{a,}$\footnote{Corresponding author: zhaowz@cnu.edu.cn}} \vskip .2in
$^a${\em School of Mathematical Sciences, Capital Normal University,
Beijing 100048, China} \\
$^b${\em Department of Mathematics, China University of Mining and Technology, Beijing 100083, China}\\
			
\begin{abstract}
We construct the generalized $\beta$ and $(q,t)$-deformed partition functions through $W$ representations,
where the expansions are respectively with respect to the generalized Jack and Macdonald polynomials labeled
by $N$-tuple of Young diagrams. We find that there are the profound interrelations between our deformed partition
functions and the $4d$ and $5d$ Nekrasov partition functions. Since the corresponding Nekrasov partition functions
can be given by vertex operators, the remarkable connection between our $\beta$ and $(q,t)$-deformed $W$-operators
and vertex operators is revealed in this paper.
In addition, we investigate the higher Hamiltonians for the generalized Jack and Macdonald polynomials.

\end{abstract}
\end{center}
		
{\small Keywords: Matrix Models, Conformal and W Symmetry, Integrable Hierarchies}
		
\end{titlepage}
	
\section{Introduction}

Nekrasov functions have attracted much attention, which are the generalizations of hypergeometric series originally introduced
to regularize the integrals over instanton moduli space \cite{Nekrasov,Nekrasov2}. The exact partition functions were obtained
as a perturbative sum over instanton sectors of coupled contour integrals. The evaluation of these integrals by the Cauchy theorem
produces a sum over residues in one-to-one correspondence with the set of boxes of Young diagrams. Alday, Gaiotto and Tachikawa
conjectured the relations (AGT conjectures or relations) between $4d$ Nekrasov partition functions with $SU(2)$ gauge groups and
$2d$ Liouville theory \cite{AGT}. It can be extended to $5d$ and $6d$ SYM theories \cite{Awata5d,5d6d} and $SU(N)$ quiver gauge
theories \cite{SUN}. The $4d$ AGT relations with $SU(N)$ gauged group can be described by a special orthogonal basis (the AFLT basis)
of the highest weight states of the $W_N\otimes Heisenberg$ algebra \cite{AFLT,Fateev12,Zhang13,Zhang14} which is labeled by
$N$-tuple of Young diagrams. With the help of AFLT bases, the Nekrasov partition functions for $U(N)$ gauge theory with $2N$
fundamental matters were given by the two-point correlation function of a particular vertex operator \cite{Awata11,Fateev12,Fukuda20,Zhang13,Zhang14,Bourgine}.
The AFLT basis, sometimes called the fixed-point basis in the geometric representation \cite{Smir,Tsymbaliuk17}, can be constructed
by the algebraic approaches. The generalized Jack polynomials (GJP) are defined by the centrally-extended spherical degenerate
double affine Hecke algebra ($\mathbf{SH^c}$) \cite{Zhang13,Zhang14,Schiffmann13}, and the generalized Macdonald polynomials (GMP) are defined by the quantum
toroidal $\mathfrak{gl}_1$ algebra \cite{Awata11,Feigin12,Tsymbaliuk17}. The GJP and GMP can be regarded
as the $4d$ and $5d$ AFLT bases, respectively. The proofs of AGT relations which are based on the Dotsenko-Fateev (DF) integral
representation of the conformal blocks \cite{DF} have been discussed \cite{AFLT,Mir11,Morozov14,SMironov,Zenkevich15,Zenkevich16,Zhang11,Fateev12}.
By decomposing the DF integral in terms of GJP or GMP, the proof of AGT relations transforms into computing the corresponding Selberg
integral averages. Moreover, this decomposition is naturally identified with the corresponding topological string amplitude,
computed using the topological vertex technique, especially in the $5d$ case \cite{Zenkevich15,Zenkevich16,Fukuda20}.

The Nekrasov partition functions can be related to matrix models \cite{Itoyama,Tai,MM12,Nishinaka,Sulkowski,Bonelli,MirPRD}.
They emerge in matrix models as a straightforward implication of superintegrability, factorization of peculiar matrix model
averages \cite{MirPRD}. $W$-representations of matrix models realize the partition functions by acting on elementary functions
with exponents of the given $W$-operators \cite{Morozov09,Kon-Wit,BGW,Cassia,MirGKM}. They are conducive to analyze the structures
of matrix models. The Hurwitz-Kontsevich (HK) matrix model can be used to describe the Hurwitz numbers and Hodge integrals over the moduli space
of complex curves \cite{Goulden01,Al1405}. There is the $W$-representation for this superintegrable matrix model, where the $W$-operator is the Hurwitz operator.
Recently, the partition function hierarchies were presented by the expansions with respect to the Schur functions via the $W$-representations,
where the $W$-operators are given by the nested commutators in which the Hurwitz operator plays a fundamental role \cite{Rui22}.
For the negative branch of hierarchy, it gives the $\tau$-functions of the KP hierarchy.
It can be described by the two-matrix model that depends on two (infinite) sets of variables and an external matrix \cite{Alexandrov23,MironovIn,MironovSkew}.
For the $\beta$-deformed partition function hierarchies in \cite{Rui22}, their integral realizations and ward identities were presented
by means of  $\beta$-deformed Harish-Chandra-Itzykson-Zuber integral \cite{Oreshina1,Oreshina2}. Furthermore, the $W$-operators in the positive branch of
hierarchies can be related to the many-body systems \cite{MironovCom}.

Much interest has been attributed to $(q,t)$-deformed matrix models \cite{szabo1,szabo2,Mor18,Mor19,Morell,Morell2,Cassia,ZenkevichNet,AwataNet,Fan23,Bou23}.
For some $(q,t)$-deformed matrix models, the $(q,t)$-deformation was carried out by substituting the Schur functions in superintegrability
expansions by the Macdonald polynomials \cite{Mor18,Mor19,Cassia}.
Ding-Iohara-Miki algebra (DIM) or quantum toroidal $\mathfrak{gl}_1$ algebra is a double centrally expanded and double quantum parameters
deformed algebra \cite{D-I,Miki07,Feigin09} and has rich representation theories \cite{Feigin12,Feigin17,Tsymbaliuk17,Bourgine17,Awata17}.
Based on DIM, some $(q,t)$-deformed partition functions have been constructed by different approaches \cite{ZenkevichNet,AwataNet,Fan23,Bou23}.
Network matrix models are naturally built from the Seiberg-Witten integrable system. Mironov et al. worked out the connection between
a large class of network matrix models associated with toric diagrams and the DIM algebra \cite{ZenkevichNet,AwataNet}.
Recently, the $(q,t)$-deformed partition functions were presented through $W$-representations, where the $W$-operators are determined
by the cut-join rotation operator $\hat O$ \cite{Fan23}. The construction of $\hat O$ can be done in terms of a Fock representation
of DIM algebra or the Macdonald difference operators.

The (centrally-extended) elliptic Hall algebra defined as Hall algebra of elliptic curve \cite{SchiffmannHall,SchiffmannKA,Schiffmann13}
may be thought as the stable limit of spherical double affine Hecke algebra (sDAHA) and isomorphic to DIM.
These generators are $\mathbb{Z}^2$ graded so that it is easier to describe the commutation relations of generators and the $SL(2,\mathbb{Z})$
automorphism (Miki automorphism).
Motivated by the recent progress in $W$-representations, in this paper, we will construct the generalized $\beta$ and $(q,t)$-deformed
$W$-operators based on the $\mathbf{SH^c}$ and elliptic Hall algebra, respectively. Our goal is to present the generalized $\beta$ and
$(q,t)$-deformed partition functions of the expansions of the GJP and GMP via $W$-representations
and to establish the connection between them and the Nekrasov partition functions.

This paper is organized as follows.
In section 2, we first construct the $\beta$-deformed $W$-operators which are closely related to the Calogero-Sutherland (CS) models
based on the Fock representation (symmetric functions representation) of the algebra $\mathbf{SH^c}$.
Then we construct the generalized $\beta$-deformed $W$-operators from the $N$-Fock representation of the algebra $\mathbf{SH^c}$
and present the partition functions of the expansions with respect to GJP through $W$-representations.
We further point out the interrelations between our partition functions and the 4$d$ Nekrasov partition functions.
In section 3, we extend the $W$-operators and partition functions
to the generalized $(q,t)$-deformed versions with the help of the elliptic Hall algebra.
We present the higher Hamiltonians for GMP. Moreover we consider the semi-classical limits and give the higher Hamiltonians for GJP.
Finally, it is shown that there are the interrelations between the generalized $(q,t)$-deformed partition functions and the $5d$ Nekrasov
partition functions. We explore the deep connection between our deformed $W$-operators and vertex operators.
We end this paper with the conclusion in section 4.

\section{Generalized $\beta$-deformed partition functions and $4d$ Nekrasov partition functions}
\subsection{Algebra $\mathbf{SH^c}$, $\beta$-deformed Hurwitz operator and deformed CS operators}

Let us recall the algebra $\mathbf{SH^c}$ and its polynomial representation \cite{Zhang13,Schiffmann13}.
The algebra $\mathbf{SH^c}$ is defined by generators $\{D_{r,s},\mathbf{c}_s\}_{r\in\mathbb{Z},s\in\mathbb{Z}_{\ge0}}$ with
commutation relations
\begin{align}
[D_{0,l},D_{\pm1,k}]&=\pm D_{\pm1,l+k-1}, & [D_{\pm 1,1},D_{\pm l,0}]
&=\pm lD_{\pm(l+1),0}, & &[D_{0,l+1},D_{\pm k,0}]=\pm D_{\pm k,l}, \nonumber \\
[D_{0,l},D_{0,k}]&=0,  &
[D_{-1,l},D_{1,k}]&=\mathbf{E}_{l+k},    & &\mathbf{c}_k\ \text{is central}, \quad k,l\ge0, \label{SHc}
\end{align}
where $D_{0,0}=0$, $\mathbf{E}_0=\mathbf{c}_0$ and $\mathbf{E}_k$ is a nonlinear combination of $\{D_{0,l},\mathbf{c}_l\}$ as follows
\begin{eqnarray}
1+(1-\beta)\sum_{l=0}^{\infty}\mathbf{E}_ls^{l+1}=
\exp\left\{\sum_{l=0}^{\infty}(-1)^{l+1}\mathbf{c}_l\pi_{l}(s)\right\}\exp\left\{\sum_{l=0}^{\infty}D_{0,l+1}\omega_{l}(s)\right\},
\end{eqnarray}
where
\begin{align*}
\pi_{l}(s)& =s^l\phi_l(1+(1-\beta)s),  \\
\omega_{l}(s)&=\sum_{q=1,-\beta,\beta-1}s^l(\phi_l(1-qs)-\phi_l(1+qs)),  \\
\phi_{0}(s)&=-\log(s),\quad \phi_l(s)=(s^{-l}-1)/l\quad l\geq1.
\end{align*}
For convenience, we take the center charges $\mathbf{c}_l=\delta_{0,l}$ in this paper.
Thus the algebra $\mathbf{SH^c}$ is generated by the triples $(D_{0,2},D_{1,0},D_{-1,0})$ here.

The algebra $\mathbf{SH^c}$ is isomorphic to the affine Yangian of $\mathfrak{gl}_1$ ($\mathcal{Y}\{\widehat{\mathfrak{gl}_1}\}$)
in \cite{Prochazka} via the following assignments
\begin{equation}
\mathbf{E}_l =\psi_l, \qquad \qquad
D_{1,l}=e_l, \qquad \qquad
D_{-1,l}=-f_l,
\end{equation}
and the specialization of $\mathbf{SH^c}$ at $\beta=1$ is isomorphic to the universal enveloping algebra of the Witt algebra $w_{1+\infty}$.

Similar to the  $\mathcal{Y}\{\widehat{\mathfrak{gl}_1}\}$, $\mathbf{SH^c}$ admits the following spectral shift automorphism
parametrized by $a\in\mathbb{C}$
\begin{eqnarray}
D_{0,s}\rightarrow\sum_{k=1}^{s}\binom{s-1}{k-1}a^{s-k}D_{0,k},\qquad \qquad D_{r,0}\rightarrow D_{r,0}.
\end{eqnarray}
Therefore the free boson representations of these generators $\{D_{0,1},D_{0,2},D_{\pm l,0}\}$ for $l\ge1$ can be defined
by power sum variables
\begin{eqnarray}
\rho_a(D_{0,2})&=&\frac{1}{2}\sum_{n,m=1}^{\infty}\left[ \beta(n+m)p_np_m\frac{\partial}
{\partial p_{n+m}} +nm p_{n+m}\frac{\partial^2}{\partial p_{n}\partial p_{m}}\right]\nonumber\\
&&+\frac{1}{2}(1-\beta)\sum_{n=1}^{\infty}(n-1)np_n\frac{\partial} {\partial p_n}+a\rho_a(D_{0,1}),\nonumber \\
\rho_a(D_{l,0})&=&p_l, \qquad   \rho_a(D_{-l,0})=\beta^{l-1}l\frac{\partial}{\partial p_l},\nonumber \\
\rho_a(D_{0,1})&=&\sum_{n=1}^{\infty}np_n\frac{\partial} {\partial p_n},
\end{eqnarray}
The actions of the operators $\rho_a(D_{0,l})$ on Jack functions are
\begin{align}
\rho_a(D_{0,l})\cdot\mathsf{Jack}_\lambda\{p_n\} =\sum_{(i,j)\in\lambda}(a+(j-1)
+\beta(1-i))^{l-1}\cdot\mathsf{Jack}_\lambda\{p_n\}, \qquad l\ge1.
\end{align}

The operator $\rho_0(D_{0,2})$ is closely related to the trigonometric Calogero-Sutherland (tCS) operator as follows:
\begin{eqnarray}
\mathcal{H}^{tr}&=&\sum_{i=1}^{L}\left(x_i\frac{\partial}{\partial x_i}\right)^2+\beta\sum_{1\le i<j\le L}\frac{x_i+x_j}
{x_i-x_j}\left(x_i\frac{\partial}{\partial x_i}-x_j\frac{\partial}{\partial x_j}\right)\nonumber \\
&=&\rho_0(D_{0,2})\left|_{p_k=\sum_{i=1}^{L}x_i^k}\right. + \rho_0(D_{0,2})\left|_{p_k=\sum_{i=1}^{L}x_i^{-k}}\right. .
\end{eqnarray}
There are also the geometric interpretations for $\rho_0(D_{0,2})$ \cite{Tsymbaliuk17,Nakajima,Lehn}.
A non-linear relation between higher Hamiltionians of tCS system and operators $\rho_a(D_{0,k})$
with $k\ge 0$ was discussed in Ref.\cite{Schiffmann13}.

In terms of the operator $\rho_0(D_{0,2})$, the $\beta$-deformed Hurwitz operator \cite{Rui22} can be expressed as
\begin{eqnarray}
W_0=\rho_{L\beta}(D_{0,2}),
\end{eqnarray}
where $L$ is the order of integrated Hermitian matrices. 	

With the help of the commutation relations (\ref{SHc}), we construct the operators $W_{\pm n}(a;p)$
by the nested commutators
\begin{eqnarray}
W_n(a;p)&=&\begin{cases}
\frac{1}{n-1}[W_{n-1}(a;p),[W_1(a;p),\rho_{\bar{a}}(D_{0,2})]], & n>1,\\
\beta^{-1}[\rho_a(D_{-1,0}),\rho_a(D_{0,2})],& n=1,\label{Wpos}
\end{cases}\nonumber \\
W_{-n}(a;p)&=&\begin{cases}
\frac{1}{n-1}[W_{1-n}(a;p),[W_{-1}(a;p),\rho_{\bar{a}}(D_{0,2})]], & n>1,\\
[\rho_a(D_{0,2}),\rho_a(D_{1,0})],& n=1,
\end{cases}\label{Wpmn}
\end{eqnarray}
where the operators $W_n(a;p)$ in the positive branch are conjugate to $W_{-n}(a;p)$
for the scaling product $\langle\cdot,\cdot\rangle_{\beta}$,
i.e., $\langle W_n f,g\rangle_{\beta}=\langle f,W_{-n}g\rangle_{\beta}$ for any symmetric functions $f$ and $g$.
We should remark that $W_{\pm n}(a;p)$ depend on the spectral parameter $a$ and $[W_{\pm n}(a;p),W_{\pm m}(a;p)]=0$ for $n,m\ge1$.

Let us list several operators
\begin{eqnarray}
W_{-1}(a;p)&=&\sum_{n=1}^{\infty}n p_{n+1}\frac{\partial}{\partial p_n}+ap_1,\nonumber\\
W_{-2}(a;p)&=&\frac{1}{2}\sum_{n,m=1}^{\infty}\left[ \beta(n+m-2)p_np_m\frac{\partial}{\partial p_{n+m-2}}
+nm p_{n+m+2}\frac{\partial^2}{\partial p_{n}\partial p_{m}}\right]\nonumber\\
&&+\frac{1}{2}(1-\beta)\sum_{n=1}^{\infty}(n+1)np_{n+2}\frac{\partial}{\partial p_n}+a\sum_{n=1}^{\infty} np_{n+2} \frac{\partial}{\partial p_n}\nonumber\\
&&+a\beta^{-1}(a-\beta+1)p_2+{a\beta^{-1}\over2}p_1^2, \nonumber\\
W_{1}(a;p)&=&\sum_{n=1}^{\infty}(n+1) p_{n}\frac{\partial}{\partial p_{n+1}}+a\beta^{-1}
\frac{\partial}{\partial p_1},\nonumber\\
W_{2}(a;p)&=&\frac{1}{2}\sum_{n,m=1}^{\infty}\left[ \beta(n+m+2)p_np_m\frac{\partial}{\partial p_{n+m+2}}
+nm p_{n+m-2}\frac{\partial^2}{\partial p_{n}\partial p_{m}}\right]\nonumber\\
&&+\frac{1}{2}(1-\beta)\sum_{n=1}^{\infty}(n+1)(n+2)p_n\frac{\partial}{\partial p_{n+2}}
+a\sum_{n=1}^{\infty}(n+2)p_n\frac{\partial}{\partial p_{n+2}}\nonumber\\
&&+2a\beta^{-2}(a-\beta+1)\frac{\partial}{\partial p_2}+{a\beta^{-3}\over2}\frac{\partial}{\partial p_1^2}.
\end{eqnarray}

The operators $W_{n}(a;p)$ can be related to the many-body systems.

(i) Let us take $a=L\beta$, $p_k=\sum_{i=1}^{L}x_i^k$ and denote $\hat{H}_n=\beta^{n-1}W_n(a;p)$ for $n\le L$,
then we obtain the rational Calogero-Sutherland (rCS) operators derived in Ref.\cite{MironovCom}
\begin{eqnarray}
\hat{H}_1&=&\sum_{i=1}^{L}\frac{\partial}{\partial x_i}, \nonumber \\
\hat{H}_2&=&\sum_{i=1}^{L}\frac{\partial^2}{\partial x_i^2}+2\beta\sum_{i\neq j}
\frac{1}{x_i-x_j}\frac{\partial}{\partial x_i}, \nonumber \\
\vdots&& \nonumber \\
\hat{H}_n&=&\sum_{k=1}^{n}\binom{n}{k}\beta^{n-k}\sum_{i=1}^{L}\left(\sum_{\overset{I\subseteq\{1,\cdots,N\}\setminus i}
{|I|=n-k}}\prod_{j\in I}\frac{1}{x_i-x_j}\right)\frac{\partial^k}{\partial x_i^k}.
\end{eqnarray}

Taking the similarity transformation
\begin{eqnarray}
\hat{\mathscr{H}}_n=\Delta^{\beta}(x)\circ\hat{H}_n\circ\Delta^{-\beta}(x)
\end{eqnarray}
with $\Delta(x)=\prod_{i<j}^{L}(x_i-x_j)$, it gives the higher Hamiltonians of the rCS
model with $L$ particles.
	
(ii) Let us take $a=L\beta-M$, $\bar{p}_k=\sum_{i=1}^{L}x_i^k-\beta^{-1}\sum_{j=1}^{M}y_j^k$ and denote
$\bar{H}_n=\beta^{n-1}W_{n}(a;\bar{p})$ for $n\le\operatorname{min}\{L,M\}$, then we obtain the deformed rCS operators
\begin{eqnarray}
\bar{H}_{1}&=&\quad\sum_{i=1}^L\frac\partial{\partial x_i}+\sum_{j=1}^M\frac\partial{\partial y_j},\nonumber\\
\bar{H}_{2}&=&\quad\sum_{i=1}^L\frac{\partial^2}{\partial x_i^2}-\beta\sum_{i=1}^M
\frac{\partial^2}{\partial y_i^2}+2\beta\sum_{i\neq j}^L\frac1{x_i-x_j}\frac\partial{\partial x_i}-2
\sum_{i\neq j}^M\frac1{y_i-y_j}\frac\partial{\partial y_i}\nonumber\\
&&-\sum_{i=1}^L\sum_{j=1}^M\frac2{x_i-y_j}\left(\frac\partial{\partial x_i}+\beta\frac\partial{\partial y_j}\right),\\
\vdots.\nonumber
\end{eqnarray}

The higher Hamiltonians of $U(L|M)$-deformed rCS model \cite{Sergeev05} are given by the similar transformation
\begin{eqnarray}
\bar{\mathscr{H} }_n=\Delta_{\beta}(x;y)\circ\bar{H}_n\circ\Delta_{\beta}^{-1}(x;y) \label{Hbar}
\end{eqnarray}
with
\begin{eqnarray}
\Delta_{\beta}(x;y)&=&\frac{\prod_{1\leq i<i'\leq L}|x_{i}-x_{i'}|^{\beta}
\prod_{1\leq j<j'\leq M}|y_{j}-y_{j'}|^{1/\beta}}{\prod_{i=1}^{L}\prod_{j=1}^{M}|x_{i}-y_{j}|}.
\end{eqnarray}
$\bar{\mathscr{H}}_1$ and $\bar{\mathscr{H}}_2$ are the momentum and Hamiltonian, respectively,
\begin{eqnarray}
\bar{\mathscr{H}}_1&=&\sum_{i=1}^{L}\frac{\partial}{\partial x_i}+\sum_{j=1}^{M}\frac{\partial}{\partial y_j},\nonumber \\
\bar{\mathscr{H}}_2&=&2(1-\beta)\left(\sum_{i<j}^L \frac{\beta}{\left(x_i-x_j\right)^2}+\sum_{i<j}^M \frac{\beta^{-1}}
{\left(y_i-y_j\right)^2}-\sum_{i=1}^L \sum_{j=1}^M\frac{1}{\left(x_i-y_j\right)^2}\right)\nonumber\\
&&+\sum_{i=1}^L \frac{\partial^2}{\partial x_i^2}-\beta\sum_{i=1}^M \frac{\partial^2}{\partial y_i^2}.
\end{eqnarray}

By means of the operators (\ref{Wpmn}), we give the partition functions through $W$-representations
\begin{subequations}
\begin{align}
Z_{-n}(a;p)&=e^{W_{-n}(a;p)}\cdot 1 \nonumber \\
&=\sum_{\lambda }\frac{\mathsf{Jack}_\lambda\{p_k=\beta^{-1}a\}}{\mathsf{Jack}_\lambda\{p_k=\beta^{-1}\delta_{k,1}\}}
\frac{\mathsf{Jack}_\lambda\{g_k=\beta^{-1}\delta_{k,n}\}\mathsf{Jack}_\lambda\{p\}}
{\langle\mathsf{Jack}_\lambda,\mathsf{Jack}_\lambda\rangle_\beta},
\label{Neg1}\\
Z_{n}(a;p|g)&=e^{W_{n}(a;p)}\cdot\exp\left\{\beta\sum_{k=1}^{\infty}\frac{1}{k}p_kg_k\right\} \nonumber \\
&=\sum_{\mu\subset\lambda}\frac{\mathsf{Jack}_\lambda\{p_k=\beta^{-1}a\}\mathsf{Jack}_\mu\{p_k=\beta^{-1}
\delta_{k,1}\}}{\mathsf{Jack}_\mu\{p_k=\beta^{-1}a\}\mathsf{Jack}_\lambda\{p_k=\beta^{-1}\delta_{k,1}\}}
\mathsf{Jack}_{\lambda/\mu}\{g_k=\beta^{-1}\delta_{k,n}\}\nonumber\\ &\cdot\frac{\mathsf{Jack}_{\mu}\{p\}
\mathsf{Jack}_{\lambda/\mu}\{g\}}{\langle \mathsf{Jack}_\lambda,\mathsf{Jack}_\lambda\rangle_\beta}\label{Pos1},
\end{align}
\end{subequations}
where $\mathsf{Jack}_{\lambda}$ is the integral form of Jack polynomial \cite{Macdonald}.

When setting $a=N\beta$ and $p_k=\sum_{i=1}^{N}x_i^k$, $Z_{-n}(a;p)$ and $Z_{n}(a;p|g)$ recover the negative
and positive branch of $\beta$-deformed partition functions in Ref.\cite{Rui22}.
Recently, their integral realizations and ward identities were presented \cite{Oreshina1,Oreshina2}.

Let us consider the case of $a=L\beta-M$ and $p_k=\sum_{i=1}^{L}x_i^k-\beta^{-1}\sum_{j=1}^{M}y_j^k$ in $Z_{\pm n}(a;p)$.
There is the super version of the $\beta$-deformed Cauchy identity
\begin{eqnarray}
\exp\left\{\beta\sum_{k=1}^{\infty}\frac{1}{k}\left(\sum_{i=1}^{L}x_i^k-\beta^{-1}\sum_{j=1}^{M}y_j^k\right)g_k\right\}=
\sum_{\lambda\in\mathcal{SP}}\frac{\mathsf{SJ}_\lambda\{x;y\}\mathsf{Jack}_\lambda\{g\}}
{\langle \mathsf{Jack}_\lambda,\mathsf{Jack}_\lambda\rangle_\beta},
\end{eqnarray}
where $\mathsf{SJ}_\lambda\{x;y\}:=\mathsf{Jack}_{\lambda}\{p\}$ is the super Jack polynomial and $\mathcal{SP}$ is
the set of the fat $(L,M)$-hook Young diagrams \cite{Sergeev05}.
By replacing $\mathsf{Jack}_{\lambda}\{p\}$ in (\ref{Neg1}) and (\ref{Pos1}) with $\mathsf{SJ}_\lambda\{x;y\}$ and restricting
the sum range of Young diagrams $\lambda$ and $\mu$ to $\mathcal{SP}$, it gives the $\beta$-deformed partition functions \cite{Fuhao}.
	
The $\beta$-deformed ABJ-like model is given by \cite{Cassia21}
\begin{eqnarray}
\tau_{L,M}(\bar{p})&=&{\frac{1}{L!}}{\frac{1}{M!}}\int_{\mathbb{R}^{L}}\prod_{i=1}^{L}\operatorname{d}x_{i}
\int_{\mathbb{R}^{M}}\prod_{j=1}^{M}\operatorname{d}y_{j}\Delta^2_{\beta}(x;y) \nonumber\\
&&\cdot\exp\left(\sum_{s=1}^{\infty}
\frac{1}{s}(\bar{p}_{s}-\delta_{s,2})\left(\sum_{i=1}^{L}x_{i}^{s}-\frac{1}{\beta}\sum_{j=1}^{M}y_{j}^{s}\right)\right).\label{ABJ}
\end{eqnarray}
It can be interpreted as the matrix model corresponding to an integral over supermatrices in the algebra of the supergroup $U(L|M)$.

It is known that the Virasoro constraints of (\ref{ABJ}) are identical to that ones of the $\beta$-deformed Gaussian-Hermitian model
just by replacing the latter's rank $L$ with the effective rank $L_{eff}=L-\beta^{-1}M$.
Thus there is a constraint equation
\begin{eqnarray}
&&\left(\sum_{s=1}^{\infty}\bar{p}_s\frac{\partial}{\partial \bar{p}_s}-2W_{-2}(\bar{p})\right)
\tau_{L,M}(\bar{p})=0. \label{constriants}
\end{eqnarray}
From (\ref{constriants}), we can confirm that the $\beta$-deformed ABJ-like model (\ref{ABJ}) is the integral realization
of $Z_{-2}(L_{eff};\bar{p})$ in (\ref{Neg1})
\begin{eqnarray}
\tau_{L,M}(\bar{p})=c_{\emptyset}e^{W_{-2}(L_{eff};\bar{p})}\cdot 1=c_{\emptyset}Z_{-2}(\beta L_{eff};\bar{p}),
\end{eqnarray}
where $c_{\emptyset}=\tau_{L,M}|_{\bar{p}=0}$.

\subsection{Generalized $\beta$-deformed partition functions}\label{GJPsection}

The algebra $\mathbf{SH^c}$ is equipped with the topological
coproduct \cite{Schiffmann13}
\begin{eqnarray}
\mathbf{\Delta}(D_{l,0})&=&\mathbf{1}\otimes D_{l,0}+D_{l,0}\otimes\mathbf{1}, \qquad l\neq0,\nonumber\\
\mathbf{\Delta}(D_{0,1})&=&\mathbf{1}\otimes D_{0,1}+D_{0,1}\otimes\mathbf{1},\\
\mathbf{\Delta}(D_{0,2})&=&\mathbf{1}\otimes D_{0,2}+D_{0,2}\otimes\mathbf{1}+(1-\beta)\sum_{l\ge1}
\beta^{1-l}lD_{-l,0}\otimes D_{l,0}\nonumber,
\end{eqnarray}
and
\begin{eqnarray}\label{coproduct}
\mathbf{\Delta}^{(1)}:&=&\mathbf{1},\nonumber\\
\mathbf{\Delta}^{(N)}:&=&(\overbrace{\mathbf{1}\otimes\cdots\otimes\mathbf{1}}^{N-2}\otimes\mathbf{\Delta})\circ\cdots\circ
(\mathbf{1}\otimes\mathbf{\Delta})\circ\mathbf{\Delta}, \qquad N\geq2.
\end{eqnarray}
Note that there should exist the coproduct structure for $D_{0,l}$, $l\geq 3$, however it is too complicated to give an explicit formula.

Let $\vec{a}=(a_1,\cdots,a_N)$ and $\vec{p}=(p^{(1)},\cdots,p^{(N)})$ be the $N$-tuples of complex parameters and time variables, respectively,
and $\rho_{\vec{a}}^{N}=\prod_{i=1}^{N}\otimes\rho_{a_i}\circ\mathbf{\Delta}^{(N-1)}$.
Then there are the $N$-Fock representation for $\mathbf{SH^c}$
\begin{eqnarray}
\rho^{N}_{\vec{a}}(D_{l,0})&=&\sum_{k=1}^{N}p_l^{(k)},  \qquad \qquad \rho^{N}_{\vec{a}}(D_{-l,0})=
\beta^{l-1}\sum_{k=1}^{N}l\frac{\partial}{\partial p_l^{(k)}}, \qquad l>0,\nonumber \\
\rho^{N}_{\vec{a}}(D_{0,1})&=&\sum_{k=1}^{N}\sum_{n\ge1}np_n^{(k)}\frac{\partial}{\partial p_n^{(k)}},\nonumber \\
\rho^{N}_{\vec{a}}(D_{0,2})&=&(1-\beta)\sum_{1\le k<l\le N}\sum_{n\ge1}n^2 p_n^{(l)}
\frac{\partial}{\partial p_n^{(k)}}+\sum_{k=1}^{N}a_k\sum_{n\ge1}np_n^{(k)}\frac{\partial}{\partial p_n^{(k)}}\nonumber\\
&&+\frac{1}{2}\sum_{k=1}^{N}\sum_{n,m\ge1}[ \beta(n+m)p_n^{(k)}p_m^{(k)}\frac{\partial}{\partial p_{n+m}^{(k)}}
+nm p_{n+m}^{(k)}\frac{\partial^2}{\partial p_{n}^{(k)}\partial p_{m}^{(k)}}\nonumber\\
&&+(1-\beta)(n-1)np_n^{(k)}\frac{\partial} {\partial p_n^{(k)}}],
\end{eqnarray}
where $p_n^{(k)}=\mathbf{1}\otimes\cdots\otimes\mathbf{1}\otimes \overset{k\mathit{-th}}{\overbrace{p_n}}
\otimes\mathbf{1}\otimes \cdots\otimes\mathbf{1}$ with $k=1,\cdots,N.$
	
Let us denote $\mathbb{W}(\vec{a})=\rho_{\vec{a}}^{N}(D_{0,2})$ and $\vec{\lambda}=(\lambda^1,\cdots,\lambda^N)$ as an $N$-tuple of Young diagrams.
The GJP $J_{\vec{\lambda}}\{\vec{a};\vec{p}\}$ are defined as
\begin{subequations}
\begin{align}
&J_{\vec{\lambda}}\{\vec{a};\vec{p}\}=(-\beta^2)^{|\vec{\lambda}|/2}\prod_{1\le k<l\le N}g_{k,l}(\vec{\lambda};\vec{a})
\tilde{J}_{\vec{\lambda}}\{\vec{a};\vec{p}\},\\
&\tilde{J}_{\vec{\lambda}}\{\vec{a};\vec{p}\} =\prod_{k=1}^{N}\mathsf{Jack}_{\lambda^k}\{p^{(k)}\}+\sum_{\vec{\mu}{<^R}\vec{\lambda}}
v^{\vec{\mu}}_{\vec{\lambda}}(\vec{a})\prod_{k=1}^{N}\mathsf{Jack}_{\mu^k}\{p^{(k)}\},\\
&\mathbb{W}(\vec{a})\cdot J_{\vec{\lambda}}\{\vec{a};\vec{p}\}=\sum_{k=1}^{N}\sum_{(i,j)\in\lambda^k}(a_k+(j-1)+(1-i)\beta)
 J_{\vec{\lambda}}\{\vec{a};\vec{p}\},
\end{align}
\end{subequations}
where $v^{\vec{\mu}}_{\vec{\lambda}}(\vec{a})$ is a complex matrix, $``{<^R}"$ is
the partial ordering on the set of $N$-tuple of Young diagrams \cite{Awata11} and
\begin{align}
&g_{k,l}(\vec{\lambda};\vec{a})=g_{\lambda^{k},\lambda^{l}}(a_k-a_l),\qquad 1\le k,l\le N,\nonumber\\
&g_{\lambda,\mu}(x)= \prod_{(i,j)\in\lambda}
[x+(\lambda_i-j+1)+\beta(\mu'_j-i)]\times\prod_{(i,j)\in\mu}[x-(\mu_i-j)-\beta(\lambda'_j-i+1)],\nonumber\\
&\langle\mathsf{Jack}_\lambda,\mathsf{Jack}_\lambda\rangle_\beta=(-\beta^{-2})^{|\lambda|}g_{\lambda,\lambda}(0),
\end{align}
especially, we have
\begin{eqnarray}
g_{\lambda,\emptyset}(x)&=&\prod_{(i,j)\in\lambda}(x+j-\beta i),\nonumber\\
g_{\emptyset,\lambda}(x)&=&\prod_{(i,j)\in\lambda}(x-(j-1)-\beta(1-i)). \label{hook2}
\end{eqnarray}
		
Since the non-symmetric term $\sum_{k<l}np_n^{(l)}\frac{\partial}{\partial p_n^{(k)}}$ in $W(\vec{a};\vec{p})$ is conjugate to
$\sum_{k>l}np_n^{(l)}\frac{\partial}{\partial p_n^{(k)}}$,
it is natural to define the dual polynomials $J^*_{\vec{\lambda}}$ by reversing the order of time variables
\begin{eqnarray}
J^*_{\vec{\lambda}}\{\vec{a};\vec{p}\}&=&J^*_{\lambda^1,\lambda^2,\cdots,\lambda^N} \{a_1,a_2,
\cdots,a_N; p^{(1)},p^{(2)},\cdots,p^{(N)}\}, \nonumber \\
&=&J_{\lambda^N,\lambda^{N-1},\cdots,\lambda^1}\{a_N,a_{N-1},\cdots,a_1; p^{(N)},p^{(N-1)},\cdots,p^{(1)}\}.
\end{eqnarray}
Then we have
\begin{eqnarray}
\langle J^*_{\vec{\mu}}\{\vec{a};\vec{p}\},J_{\vec{\lambda}}\{\vec{a};\vec{p}\}\rangle_\beta=
\prod_{k,l=1}^{N}g_{k,l}(\vec{\lambda};\vec{a})\delta_{\vec{\mu},\vec{\lambda}}.
\end{eqnarray}
It implies the Cauchy completeness identity for GJP
\begin{equation}
\sum_{\vec{\lambda}}\frac{J_{\vec{\lambda}}\{\vec{a};\vec{p}\}J^*_{\vec{\lambda}}\{\vec{a};\vec{g}\}}
{\prod_{k,l=1 }^{N}g_{k,l}(\vec{\lambda};\vec{a})}z^{|\vec{\lambda}|}=\exp\left\{\beta\sum_{k=1}^{N}\sum_{n\ge1}
\frac{p_n^{(k)}g_n^{(k)}}{n}z^n\right\}.
\end{equation}
		
Let us introduce the operators $\mathbb{E}_n$ and $\mathbb{F}_n$ for $n\ge1$ as follows:
\begin{eqnarray}
\mathbb{E}_n&=&\mathbb{E}_n(\vec{a};\vec{m})=\operatorname{ad}_{\mathbb{W}(\vec{a}^{n})}\cdots\mathrm{ad}_{\mathbb{W}
(\vec{a}^{1})}\mathbb{E}_0,  \nonumber \\
\mathbb{F}_n&=&\mathbb{F}_n(\vec{a};\vec{m})=(-1)^n\mathrm{ad}_{\mathbb{W}(\vec{a}^{n})}\cdots\mathrm{ad}_{\mathbb{W}
(\vec{a}^{1})}\mathbb{F}_0,
\end{eqnarray}
where
\begin{align}
\mathbb{E}_0=\rho^N_{\vec{a}}(D_{1,0}), \qquad
\mathbb{F}_0=\beta^{-1}\rho^N_{\vec{a}}(D_{-1,0}),
\end{align}
$\mathrm{ad}_f\ g:=[f,g]$, $\vec{m}=(m_1,m_2,\cdots)$ and $\vec{a}^{r}=(a_1+m_r,\cdots,a_N+m_r)$.

In terms of $\mathbb{E}_n$ and $\mathbb{F}_n$, we construct the generalized $\beta$-deformed $W$-operators with $N$-tuple of time variables as follows:
\begin{subequations}\label{betaW}
\begin{eqnarray}
\sum_{l\ge1}W_{-l,nl}(\vec{a};\vec{m})s^{l-1} &=&e^{s\cdot \mathrm{ad}_{\mathbb{E}_{n+1}}}(\mathbb{E}_{n}),\qquad n\ge1,\label{W-} \\
\sum_{l\ge1}W_{l,nl}(\vec{a};\vec{m})s^{l-1} &=&e^{-s\cdot \mathrm{ad}_{\mathbb{F}_{n+1}}}(\mathbb{F}_{n}),\qquad n\ge1,\\
\sum_{n=0}^{\infty}W_{0,n+1}(\vec{a};m) \omega_{n}(s)&=& \log\left(1+\frac{1-\beta}{1+(1-\beta)s}\sum_{n=1}^{\infty}
K_n(\vec{a};m)s^{n+1}\right)\label{Woper0},
\end{eqnarray}
\end{subequations}
where
\begin{eqnarray*}
K_n(\vec{a})\equiv[\mathbb{E}_r ,\mathbb{F}_{n-r}] |_{m_i\equiv m},\quad 0\le r\le n.
\end{eqnarray*}
It is clear that $\mathbb{E}_n=W_{-1,n}(\vec{a};\vec{m})$, $\mathbb{F}_n=W_{1,n}(\vec{a};\vec{m})$, $\mathbb{W}(\vec{a}^r)=W_{0,2}(\vec{a}^r)$
for $r\ge1$ and $W_{l,nl}(\vec{a};\vec{m})$ are conjugate to $W_{-l,nl}(\vec{a};\vec{m})$.

When restricted to the $N=1$ case, the operators $W_{\pm l,nl}(\vec{a};\vec{m})$ degenerate to $H^{(n)}_{\pm l}$
in Ref.\cite{MironovSkew} which are labeled by two indices, $l$ is called the grading and $nl+1$ is the spin.
A noteworthy property is that the $H^{(n)}_{\pm l}$ can be embedded in the $W_{1+\infty}$ algebra.
	
We observe that the explicit computation of actions of the operators $W_{\pm l,nl}(\vec{a};\vec{m})$ (\ref{betaW}) on GJP become rather cumbersome.
To overcome this problem, we make use of the technique of cut-and-join rotation operators \cite{Al1405,MironovIn,MironovCom,Fan23}.
Let us introduce the generalized cut-and-join rotation operator
\begin{eqnarray}
\hat{O}_{\beta}(\vec{a};x;\vec{p})=\exp\left\{\rho_{\vec{a}} (D_{0,1})\cdot\log x+\sum_{n=1}^{\infty}
\frac{(-x^{-1})^n}{n}\rho_{\vec{a}}(D_{0,n+1})\right\},
\end{eqnarray}
such that
\begin{eqnarray}
\hat{O}_{\beta}(\vec{a};x;\vec{p})\cdot J_{\vec{\lambda}}\{\vec{a};\vec{p}\}=\prod_{k=1}^{N}
g_{\lambda^k,\emptyset}(x+a_k-1+\beta) J_{\vec{\lambda}}\{\vec{a};\vec{p}\}.
\end{eqnarray}

With the help of the recursive formulas
\begin{align}
[W_{-1,1}(\vec{a};\vec{m};\vec{p}),W_{-l,0}(\vec{a};\vec{m};\vec{p})]&=lW_{-(l+1),0}(\vec{a};\vec{m};\vec{p}), \nonumber \\
[W_{0,2}(\vec{a}^{r};\vec{p}),W_{-1,r-1}(\vec{a};\vec{m};\vec{p})]&=\hat{O}_{\beta}(\vec{a};m_{r};\vec{p})
\circ W_{-1,r-1}(\vec{a};\vec{m};\vec{p})\circ\hat{O}_{\beta}^{-1}(\vec{a};m_{r};\vec{p}),
\end{align}
and the similar relations for $W_{-l,nl}(\vec{a};\vec{m};\vec{p})$, the $W$-operators (\ref{betaW}) can be expressed as
\begin{eqnarray}
W_{-l,nl}(\vec{a};\vec{m};\vec{p})&=&\prod_{i=1}^{n}\hat{O}_{\beta}(\vec{a};m_i;\vec{p})\circ \sum_{k=1}^{N}p_l^{(k)}
\circ \prod_{i=1}^{n}\hat{O}^{-1}_{\beta}(\vec{a};m_i;\vec{p})\nonumber\\
&=&\sum_{k=1}^{N}\hat{W}^{(k)}_{-l,nl}(\vec{a};\vec{m};\vec{p}),\nonumber\\
W_{l,nl}(\vec{a};\vec{m};\vec{p})&=&\prod_{i=1}^{n}\hat{O}^{-1}_{\beta}(\vec{a};m_i;\vec{p})\circ \beta^{-1}\sum_{k=1}^{N}l
\frac{\partial}{\partial{p_l}^{(k)}}\circ \prod_{i=1}^{n}\hat{O}_{\beta}(\vec{a};m_i;\vec{p})\nonumber\\
&=&\sum_{k=1}^{N}\hat{W}^{(k)}_{l,nl}(\vec{a};\vec{m};\vec{p}) \label{Woper}.
\end{eqnarray}
The actions of the operators (\ref{Woper}) and $W_{0,n}(\vec{a})$ in (\ref{Woper0}) on GJP are
\begin{subequations}
\begin{eqnarray}
W_{-l,nl}(\vec{a};\vec{m};\vec{p}) J_{\vec{\lambda}}\{\vec{a};\vec{p}\}&=&\sum_{|\vec{\mu}/\vec{\lambda}|=
l}C_{\vec{\lambda}}^{\vec{\mu}} \prod_{k=1}^{N}\prod_{r=1}^{n} \frac{g_{\mu^k,\emptyset}(a_k+m_r-1+\beta)}
{g_{\lambda^k,\emptyset}(a_k+m_r-1+\beta)} J_{\vec{\mu}}\{\vec{a};\vec{p}\},\\
W_{l,nl}(\vec{a};\vec{m};\vec{p})J_{\vec{\lambda}}\{\vec{a};\vec{p}\}&=&\sum_{|\vec{\lambda}/\vec{\mu}|=l}
\bar{C}_{\vec{\lambda}}^{\vec{\mu}}\prod_{k=1}^{N}\prod_{r=1}^{n} \frac{g_{\lambda^k,\emptyset}(a_k+m_r-1+\beta)}
{g_{\mu^k,\emptyset}(a_k+m_r-1+\beta)} J_{\vec{\mu}}\{\vec{a};\vec{p}\},\\
W_{0,n}(\vec{a};\vec{p})J_{\vec{\lambda}}\{\vec{a};\vec{p}\}&=&\sum_{k=1}^{N}\sum_{(i,j)\in\lambda^k}
(a_k+(j-1)+\beta(1-i))^{n-1}\cdot J_{\vec{\lambda}}\{\vec{a};\vec{p}\},\label{eigenGJP}
\end{eqnarray}
\end{subequations}
where the coefficients $C_{\vec{\lambda}}^{\vec{\mu}}$ and $\bar{C}_{\vec{\lambda}}^{\vec{\mu}}$ are given by
\begin{eqnarray}
\sum_{|\vec{\mu}/\vec{\lambda}|=l}C_{\vec{\lambda}}^{\vec{\mu}} J_{\vec{\mu}}\{\vec{a};\vec{p}\}
&=&\sum_{k=1}^{N}p_l^{(k)}\cdot J_{\vec{\lambda}}\{\vec{a};\vec{p}\},\nonumber\\
\sum_{|\vec{\lambda}/\vec{\mu}|=l}\bar{C}_{\vec{\lambda}}^{\vec{\mu}}J_{\vec{\mu}}\{\vec{a};\vec{p}\}
&=&\beta^{-1}\sum_{k=1}^{N}l\frac{\partial}{\partial p_l^{(k)}}\cdot J_{\vec{\lambda}}\{\vec{a};\vec{p}\}.
\end{eqnarray}
	
Due to the desired $W$-operators, we construct the generalized $\beta$-deformed partition functions
\begin{subequations}
\begin{align}
Z_{-,n}\{z;\vec{p};\vec{g}\}&=\exp\left\{\beta\sum_{l\ge1}\frac{z^l}{l}\sum_{k=1}^{N}W^{(k)}_{-l,nl}
(\vec{a};\vec{m};\vec{p})W^{(k)}_{-l,nl}(\vec{a};\vec{m}';\vec{g})\right\}\cdot1 \nonumber \\
&=\sum_{\vec{\lambda}}z^{|\vec{\lambda}|}\prod_{k=1}^{N}\prod_{r=1}^{n}g_{\lambda^k,\emptyset}(a_k+m_r-1+\beta)
(-1)^{|\lambda^k|}g_{\emptyset,\lambda^k}(-a_k-m'_r)\nonumber\\
&\cdot\frac{J^*_{\vec{\lambda}}\{\vec{a};\vec{p}\}J_{\vec{\lambda}}\{\vec{a};\vec{g}\}}{\prod_{k,l=1}^{N}g_{k,l}
(\vec{\lambda})},\label{GJP}\\
Z_{0,n}\{z;\vec{p};\vec{g}\}&=\exp\{W_{0,n}(\vec{a};\vec{p})\}\exp\left\{\beta\sum_{l\ge1}\frac{z^l}{l}
\sum_{k=1}^{N}p_l^{(k)}g_l^{(k)}\right\} \nonumber \\
&=\sum_{\vec{\lambda}}z^{|\vec{\lambda}|}\exp\left\{\sum_{k=1}^{N}\sum_{(i,j)\in\lambda^k}(a_k+(j-1)
+\beta(1-i))^{n-1}\right\}\nonumber\\
&\cdot\frac{J^*_{\vec{\lambda}}\{\vec{a};\vec{p}\}J_{\vec{\lambda}}\{\vec{a};\vec{g}\}}
{\prod_{k,l=1}^{N}g_{k,l}(\vec{\lambda})},\\
Z_{+,n}\{z,w;\vec{p};\vec{g}\}&=\exp\left\{\beta\sum_{l\ge1}\frac{z^l}{l}\sum_{k=1}^{N}W^{(k)}_{l,nl}
(\vec{a};\vec{m};\vec{p})W^{(k)}_{l,nl}(\vec{a};\vec{m}';\vec{g})\right\}\nonumber\\
&\cdot\exp\left\{\beta\sum_{l\ge1}
\frac{w^{-l}}{l}\sum_{k=1}^{N}p_l^{(k)}g_l^{(k)}\right\} \nonumber \\ &=\sum_{\vec{\mu}\subset\vec{\lambda}}
\frac{z^{|\vec{\mu}|}}{w^{|\vec{\lambda}|}}\prod_{k=1}^{N}\prod_{r=1}^{n}g_{\mu^k,\emptyset}
(a_k+m_r-1+\beta)(-1)^{|\mu^k|}g_{\emptyset,\mu^k}(-a_k-m'_r)\nonumber \\
&\cdot\prod_{k,l=1}^{N}\frac{g_{k,l}(\vec{\mu})}{g_{k,l}(\vec{\lambda})}J^*_{\vec{\lambda}/\vec{\mu}}
\{\vec{a};\vec{p}\}J_{\vec{\lambda}/\vec{\mu}}\{\vec{a};\vec{g}\},
\end{align}
\end{subequations}
where the skew polynomials $J_{\vec{\lambda}/\vec{\mu}}$ are defined by
\begin{eqnarray}
\langle f,J_{\vec{\lambda}/\vec{\mu}} \rangle_{\beta}&=&\frac{1}{\langle J^*_{\vec{\mu}}, J_{\vec{\mu}}\rangle_{\beta}}
\langle fJ^*_{\vec{\mu}}, J_{\vec{\lambda}}\rangle_{\beta},\nonumber\\
\langle f,J^*_{\vec{\lambda}/\vec{\mu}} \rangle_{\beta}&=&\frac{1}{\langle J^*_{\vec{\mu}}, J_{\vec{\mu}}
\rangle_{\beta}}\langle fJ_{\vec{\mu}}, J^*_{\vec{\lambda}}\rangle_{\beta},
\end{eqnarray}
for any symmetric function $f$.

\subsection{$4d$ Nekrasov partition functions}

The $4d$ Nekrasov partition function for the $U(N)$ theory with $N_f=2n$ fundamental hypermultiplets is given
by \cite{Nekrasov,Nekrasov2,AGT}
\begin{eqnarray}
Z_{inst}^{U(N),n}(z;\vec{a};\vec{m}^{+},\vec{m}^-) =\sum_{\vec{\lambda}}z^{|\vec{\lambda}|}Z_{vec}(\vec{a},\vec{\lambda})
\prod_{f=1}^{n}Z_{fund}(\vec{a},\vec{\lambda};m^+_f)Z_{anti}(\vec{a},\vec{\lambda};m^-_f),\label{4dNek}
\end{eqnarray}
where
\begin{eqnarray}
Z_{vec}(\vec{a},\vec{\lambda})&=&\prod_{k,l=1}^{N}g^{-1}_{k,l}(\vec{\lambda}),\nonumber\\
Z_{fund}(\vec{a},\vec{\lambda};m^+)&=&\prod_{k=1}^{N}\prod_{(i,j)\in\lambda^{k}}(a_k-m^++j-\beta i)=
\prod_{k=1}^{N}g_{\lambda^{k},\emptyset}(a_k-m^+),\nonumber	\\ Z_{anti}(\vec{a},\vec{\lambda};m^-)&=&\prod_{k=1}^{N}
\prod_{(i,j)\in\lambda^{k}}[a_k+m^-+(j-1)+\beta(1-i)]\nonumber\\
&=&\prod_{k=1}^{N}(-1)^{|\lambda^{k}|}g_{\emptyset,\lambda^{k}}(-a_k-m^-).
\end{eqnarray}

The partition function (\ref{4dNek}) depends on the mass vectors $\vec{m}^+$ and $\vec{m}^-$ for fundamental
and anti-fundamental respectively with $n$ components.
Note that we take the supergravity background parameters $(\epsilon_1,\epsilon_2)=(-\beta,1)$ here.
			
The symmetric function space is isomorphic to the boson Fock space by the means of
\begin{eqnarray*}
\iota: && p_{\lambda}\rightarrow |p_{\lambda}\rangle= \hat{a}_{-\lambda_1}\hat{a}_{-\lambda_2}\cdots|0\rangle,\\
\iota^*:&& p_{\lambda}\rightarrow \langle p_{\lambda}|=\langle0|\cdots \hat{a}_{\lambda_2}\hat{a}_{\lambda_1},
\end{eqnarray*}
such that
\begin{equation}
\langle p_{\lambda}|p_{\mu}\rangle=\langle0| \hat{a}_{-\lambda}\hat{a}_{\mu}|0\rangle=
\langle p_\lambda,p_\mu\rangle_\beta \label{inner},
\end{equation}
where $[\hat{a}_n,\hat{a}_m]=n\beta^{-1}\delta_{m+n,0}$.
	
For the vertex operator \cite{CO}
\begin{eqnarray}
\hat{v}(x)=\exp\left(\sum_{n=1}^{\infty}\frac{(-1)^n}{n}\hat{a}_{-n}\right)^{x-1
+\beta}\cdot\exp\left(\sum_{n=1}^{\infty}\frac{(-1)^n}{n}\hat{a}_n\right)^{-x},
\end{eqnarray}
it satisfies a Pieri-type formula
\begin{eqnarray}
\langle \mathsf{Jack}_\lambda|\hat{v}(x)| \mathsf{Jack}_\mu\rangle=
(-1)^{|\lambda|}\beta^{-|\lambda|-|\mu|}g_{\lambda,\mu}(-x).
\end{eqnarray}

For the $4d$ $U(1)$ Nekrasov partition function with $\mathcal{N}_{f}=2$ and $z=-1$, it can be
written in terms of the vertex operator as
\begin{eqnarray}
Z^{U(1),1}_{inst}(-1;0;x_1;x_2)&=&
\langle0|\hat{v}(x_2)\hat{v}(x_1)|0\rangle\nonumber\\
&=&\sum_{\lambda}\frac{\langle 0|\hat{v}(x_2)|\mathsf{Jack}_\lambda\rangle\langle
\mathsf{Jack}_\lambda|\hat{v}(x_1)|0\rangle}{\langle \mathsf{Jack}_\lambda|\mathsf{Jack}_\lambda\rangle}\nonumber\\
&=&\sum_{\lambda}\frac{g_{\lambda,\emptyset}(-x_1)g_{\emptyset,\lambda}(-x_2)}{g_{\lambda,\lambda}(0)}.
\end{eqnarray}
	
More generally, there exists the vertex operator $\hat{V}(x)=\tilde{V}_HV_W$  \cite{Fateev12,Zhang13},
where $V_W$ is the vertex operator for $W_N$ algebra and $\tilde{V}_H$ describes the contribution of $U(1)$ factor,
such that
\begin{eqnarray}
\langle J_{\vec{\mu}}(\vec{b})|\hat{V}(x)| J_{\vec{\lambda}}(\vec{a}) \rangle=\prod_{i,j=1}^{N}g_{\lambda^i,\mu^j}(a_i-b_j-x).
\end{eqnarray}

Similarly, in terms of this vertex operator $\hat{V}(x)$, one can give the $4d$ $U(N)$ Nekrasov partition
function with $\mathcal{N}_{f}=2N$ fundamental matters
\begin{eqnarray}
Z^{U(N),N}_{inst}((-1)^N;\vec{a};\vec{m}^+;\vec{m}^-)&=&\langle 0|\hat{V}^{\vec{b}}_{\vec{a}}(x_2)
\hat{V}^{\vec{a}}_{\vec{c}}(x_1) |0 \rangle \nonumber\\
&=&\sum_{\vec{\lambda}}\frac{\langle 0 |\hat{V}^{\vec{b}}(x_2)|J_{\vec{\lambda}}(\vec{a})\rangle
\langle J_{\vec{\lambda}}(\vec{a})| \hat{V}_{\vec{c}}(x_1)|0 \rangle}{\langle J_{\vec{\lambda}}
(\vec{a})|J_{\vec{\lambda}}(\vec{a})\rangle}\nonumber\\
&=&\sum_{\vec{\lambda}}\frac{\prod_{i,j=1}^{N}\left[g_{\lambda^i,\emptyset}(a_i-b_j-x_1)\cdot
g_{\emptyset,\lambda^i}(c_j-a_i-x_2)\right]}{\prod_{k,l=1}^{N}g_{k,l}(\vec{\lambda})},\label{betavertex}
\end{eqnarray}
where $m^{+}_i=x_1+b_i$, $m^{-}_i=x_2-c_i$ for $i=1,\cdots,N$.

Let us turn to our partition functions (\ref{GJP}). We observe that by removing the generalized Jack polynomials from (\ref{GJP})
and taking $m_i=1-\beta-m^{+}_i$, $m'_i=m^{-}_i$ for $i=1,2,\cdots,n$,
the remains match with the $4d$ Nekrasov partition functions  (\ref{4dNek}).
An interesting point about this observation is that now we can establish the connection between these two partition functions
\begin{eqnarray}
Z_{inst}^{U(N),n}(z;\vec{a};\vec{m}^+;\vec{m}^-)=\left\langle\sum_{\vec{\lambda}}
\frac{J^*_{\lambda}\{\vec{a};\vec{p}\}
J_{\lambda}\{\vec{a};\vec{g}\}}{\prod_{k,l=1}^{N}g^2_{k,l}(\lambda)},
Z_{-,n}(z;\vec{p};\vec{g})\right\rangle_{\beta}.\label{betavertexwope}
\end{eqnarray}

Then we make the explicit connection between $W$-operators (\ref{Woper}) and the vertex operator $\hat{V}(x)$
from (\ref{betavertex}) and (\ref{betavertexwope})
\begin{eqnarray}
&&\left\langle\sum_{\vec{\lambda}}\frac{J^*_{\lambda}\{\vec{a};\vec{p}\}
J_{\lambda}\{\vec{a};\vec{g}\}}{\prod_{k,l=1}^{N}g^2_{k,l}(\lambda)},\exp\left\{\sum_{l\ge1}A_l
\sum_{k=1}^{N}W^{(k)}_{-l,Nl}(\vec{a};\vec{m};\vec{p})W^{(k)}_{-l,Nl}(\vec{a};\vec{m}';\vec{g})\right\}
\cdot1\right\rangle_{\beta} \nonumber\\
&=&\langle 0|\hat{V}^{\vec{b}}_{\vec{a}}(x_2) \hat{V}^{\vec{a}}_{\vec{c}}(x_1) |0 \rangle,
\end{eqnarray}
where $A_l=(-1)^{Nl}\beta l^{-1}$ and $m_i=1-\beta-b_i-x_1$, $m'_i=x_2-c_i$ for $i=1,\cdots,N$.
	
Let us denote $W_{Dessin}(u,v)$ as $W_{-1,2}(\vec{a};\vec{m})$ in (\ref{W-}) with $\beta=1,(\vec{a};\vec{m})=(0;u,v)$,
then
\begin{eqnarray}
W_{Dessin}(u,v)&=&\frac{1}{2}\sum_{a,b\ge1}\left[(a+b-1)p_ap_b\frac{\partial}{\partial p_{a+b-1}}+abp_{a+b+1}
\frac{\partial^2}{\partial p_a\partial p_b}\right]\nonumber\\
&&+(u+v)\sum_{a\ge1}p_{a+1}\frac{\partial}{\partial p_{a}}+uvp_1,
\end{eqnarray}
and the corresponding partition function with $w$-representation and its character expansions are given by
\begin{eqnarray}
Z_{Dessin}\{s,u,v;p\}&=&e^{sW_{Dessin}(u,v)}\cdot1\nonumber\\ &=&\sum_{\lambda}s^{|\lambda|}\frac{\mathsf{Schur}_\lambda\{p_k=u\}
\mathsf{Schur}_\lambda\{p_k=v\}}{\mathsf{Schur}_\lambda\{p_k=\delta_{k,1}\}}\frac{\mathsf{Schur}_\lambda\{p\}}
{\langle\mathsf{Schur}_\lambda,\mathsf{Schur}_\lambda\rangle}\label{Zdessin},
\end{eqnarray}
which is nothing but the dessins d'enfant tau-function \cite{ZhouJ}.

It is known that $Z_{Dessin}=e^{\mathcal{F}_{Dessin}(s,u,v)}$, where the free energy is defined as the generating series of
weighted count of labeled dessins d'enfants
\begin{align}
&\mathcal{F}_{Dessin}(s,u,v,p_1,p_2,\ldots)\nonumber\\
=&\sum_{k,l,m\geq1}\frac1{m!}\sum_{\mu_1,...,\mu_m\geq1}N_{k,l}(\mu_1,\ldots,\mu_m)s^{|\mu|}u^kv^lp_{\mu_1}\cdots p_{\mu_m}.
\label{free}
\end{align}

Comparing (\ref{4dNek}) with (\ref{Zdessin}) and (\ref{free}), we denote
\begin{eqnarray}
\mathcal{F}_{inst}(s,u,v)&\equiv&\mathcal{F}_{Dessin}(s,u,v;p)|_{p_k=\delta_{k,1}} \nonumber\\
&=&\sum_{k,l,m\geq1}\frac1{m!}N_{k,l}(1^m)s^{m}u^kv^l,
\end{eqnarray}
then we have
\begin{eqnarray} Z_{inst}^{U(1),1}\left.\left(-s;0;-u;-v\right)\right|_{\beta=1}=e^{\mathcal{F}_{inst}(s,u,v)} \label{dessin}.
\end{eqnarray}
It means that the free energy of the $U(1)$ instanton partition function with $\mathcal{N}_f=2$ and $\beta=1$
is given by the specific part of (connected) Belyi fat graphs.

\section{Generalized $(q,t)$-deformed partition functions and $5d$ Nekrasov partition functions}
	
\subsection{Elliptic Hall algebra and Higher Hamiltonians for GMP}

Let us recall the elliptic Hall algebra $\hat{\mathcal{E}}$ \cite{SchiffmannHall,SchiffmannKA}
which is $K$-algebra generated by elements  $\{\mathbf{u}_{k,l}\}_{k,l\in\mathbb{Z}}$ and centers $(\mathbf{c}_{0,1},\mathbf{c}_{1,0})$,
modulo the following relations
\begin{subequations}
\begin{align}
[\mathbf{u}_{0,0},\mathbf{u}_{k,l}]&=k\mathbf{u}_{k,l},\nonumber\\
[\mathbf{u}_{0,k},\mathbf{u}_{1,l}]&=\operatorname{sgn}(k) \mathbf{c}_{0,1}^{(k-|k|)/2}\mathbf{u}_{1,l+k},&  &k\neq0,\\
[\mathbf{u}_{0,k},\mathbf{u}_{-1,l}]&=-\operatorname{sgn}(k) \mathbf{c}_{0,1}^{-(k+|k|)/2}\mathbf{u}_{-1,l+k},&  &k\neq0,\nonumber\\
[\mathbf{u}_{ra,rb},\mathbf{u}_{sa,sb}]&=s\frac{(\mathbf{c}_{0,1}^{b}\mathbf{c}_{1,0}^a)^{-s}-(\mathbf{c}_{0,1}^b
\mathbf{c}_{1,0}^a)^s}{\kappa_s}\delta_{r,-s},& &\operatorname{gcd}(a,b)=1,\\
[\mathbf{u}_{-1,k},\mathbf{u}_{1,l}]&=
\begin{cases}
-\kappa_1^{-1}\mathbf{c}_{0,1}^{-k}\mathbf{c}_{1,0}\theta_{0,k+l},&k+l>0,\\
\kappa_1^{-1}(\mathbf{c}_{0,1}^{k}\mathbf{c}_{1,0}{-1}-\mathbf{c}_{0,1}^{-k}\mathbf{c}_{1,0}),&k+l=0,\\
\kappa_1^{-1}\mathbf{c}_{0,1}^{-l}\mathbf{c}_{1,0}^{-1}\theta_{0,k+l},&k+l<0,
\end{cases}
\end{align}	
\end{subequations}
where $\kappa_s=(1-q^s)(1-t^{-s})(1-(t/q)^s)$ and
\begin{eqnarray}
\sum_{s=0}^{\infty}\theta_{sk,sl}z^s=\exp\left(-\sum_{r=1}^{\infty}\frac{\kappa_r}{r}\mathbf{u}_{rk,rl}z^r\right),
\end{eqnarray}
for all $(k,l)\in\mathbb{Z}^2\setminus\{0,0\}$ with $\operatorname{gcd}(k,l)=1$. The $\operatorname{gcd}(a,b)=1$
represents that the positive integers $a$ and $b$ are coprime.
	
$\hat{\mathcal{E}}$ is $\mathbb{Z}^2$ graded and has a natural $SL(2,\mathbb{Z})$ automorphism
\begin{eqnarray}
\sigma:\mathbf{u}_{k,l}\mapsto\mathbf{u}_{l,-k},\qquad \mathbf{c}_{0,1}\mapsto\mathbf{c}_{1,0},\qquad
\mathbf{c}_{1,0}\mapsto\mathbf{c}_{0,1}^{-1}.\label{automorphism}
\end{eqnarray}
It can be generated by the fundamental elements $\mathbf{u}_{0,\pm1}$, $\mathbf{u}_{\pm1,0}$, $\mathbf{c}_{0,1}$ and $\mathbf{c}_{1,0}$.

The Cartan generators $\theta_{0,\pm s}$ and $\theta_{\pm s,0}$ are generated by
\begin{subequations}
\begin{eqnarray}
\theta_{0,\pm s}&=&\oint\frac{\mathrm{d}z}{z^{s+1}} \exp\left\{-\sum_{r=1}^{\infty}\frac{\kappa_r}{r}
\mathbf{u}_{0,\pm r} z^r\right\}\nonumber\\ &=&
\begin{cases}
1, & s=0,\\
-\kappa_1\mathbf{c}_{1,0}^{\mp1} [\mathbf{u}_{\mp 1,0},\operatorname{ad}^{s}_{\mathbf{u}_{0, \pm1}}\mathbf{u}_{\pm1,0}],& s\ge1,
\end{cases}\label{theta1}\\
\theta_{\pm s,0}&=&\oint\frac{\mathrm{d}z}{z^{s+1}} \exp\left\{-\sum_{r=1}^{\infty}\frac{\kappa_r}{r}
\mathbf{u}_{\pm r,0} z^r\right\}\nonumber\\ &=&
\begin{cases}
1, & s=0,\\
-\kappa_1\mathbf{c}_{0,1}^{\pm1} [\mathbf{u}_{0,\pm1},\operatorname{ad}^{s}_{\mathbf{u}_{\pm 1,0}}\mathbf{u}_{0,\mp1}],& s\ge1.
\end{cases}\label{theta}
\end{eqnarray}
\end{subequations}
$\hat{\mathcal{E}}$ is isomorphic to the  quantum toroidal algebra of $\mathfrak{gl}_1$ in \cite{Feigin17} via the following assignments
\begin{align}
\mathbf{u}_{1,n}&\mapsto e_n,& \mathbf{u}_{-1,n}&\mapsto f_n, &  -\frac{1}{r}\mathbf{c}_{0,1}^{\mp}
\mathbf{u}_{0,\pm r}\mapsto h_{\pm r}, \nonumber\\
\mathbf{c}_{0,1}&\mapsto C,& \mathbf{c}_{1,0}&\mapsto (C^\perp)^{-1},&   n\in\mathbb{Z},\ r\ge0.
\end{align}
		
We take the centers $(\mathbf{c}_{0,1},\mathbf{c}_{1,0})=(1,(t/q)^{1/2})$ in this paper. There is the boson Fock
representation of $\hat{\mathcal{E}}$ by power sum variables \cite{SchiffmannKA,Zenkevich23} such as
\begin{subequations}
\begin{eqnarray}
\rho_u(\mathbf{u}_{k,-1})&=&\frac{u^{-1}}{(1-q^{-1})(1-t)}\oint\frac{\mathrm{d}z}{z^{k+1}}\exp\left(-\sum_{n=1}^\infty
\frac{1-t^{-n}}{n}(t/q)^{n/2}p_nz^n\right)\nonumber\\
&&\cdot\exp\left(\sum_{n=1}^\infty(1-q^n)\frac{\partial}{\partial p_n}(t/q)^{n/2}z^{-n}\right),\\
\rho_u(\mathbf{u}_{k,1})&=&-\frac{u}{(1-q)(1-t^{-1})}\oint\frac{\mathrm{d}z}{z^{k+1}}\exp\left(\sum_{n=1}^\infty
\frac{1-t^{-n}}{n}p_nz^n\right)\nonumber\\
&&\cdot\exp\left(-\sum_{n=1}^\infty(1-q^n)\frac{\partial}{\partial p_n}z^{-n}\right),\\
\rho_u(\mathbf{u}_{k,0})&=&
\begin{cases}
-\frac{(t/q)^{-k/2}}{1-q^k}p_k,& k>0,\\
\sum_{n\ge1}np_n\frac{\partial}{\partial p_n},& k=0,\\
-k\frac{1}{1-t^{k}}\frac{\partial}{\partial p_{-k}},&k<0,
\end{cases}
\end{eqnarray}
\end{subequations}
and the actions
\begin{eqnarray}
\rho_u(\mathbf{u}_{0,\pm n})\cdot\mathsf{Mac}_\lambda\{p\}=\pm u^{\pm n}\mathbf{C}_{\lambda}^{(\pm n)}
\mathsf{Mac}_\lambda\{p\}, \qquad n>0,
\end{eqnarray}
where $u$ is a nonzero complex parameter, $\mathsf{Mac}_\lambda\{p\}$ are integral form of the Macdonald polynomials and
$\mathbf{C}_{\lambda}^{(n)}=\sum_{(i,j)\in\lambda}(q^{j-1}t^{1-i})^n-\frac{1}{(1-q^n)(1-t^{-n})}$ for $n\neq0$.
		
$\hat{\mathcal{E}}$ is equipped with the topological coproduct structure \cite{Schiffmann13}
\begin{eqnarray}
\Delta(\mathbf{u}_{0,n})&=&\mathbf{u}_{0,n}\otimes\mathbf{1}+\mathbf{c}_{0,1}^n\otimes \mathbf{u}_{0,n}, \nonumber\\
\Delta(\mathbf{u}_{-1,n})&=&\mathbf{u}_{-1,n}\otimes\mathbf{1}+\sum_{k\ge0}\mathbf{c}_{0,1}^{n+k}
\mathbf{c}^{-1}_{1,0}\theta_{0,-k}\otimes \mathbf{u}_{-1,n+k},\\
\Delta(\mathbf{u}_{1,n})&=&\mathbf{u}_{1,n}\otimes\mathbf{1}+\sum_{k\ge0}\mathbf{c}_{0,1}^{n-k}\mathbf{c}_{1,0}
\theta_{0,k}\otimes \mathbf{u}_{1,n-k}.\nonumber
\end{eqnarray}
We denote $\vec{u}=(u_1,\cdots,u_n)$ as an $N$-tuple of nonzero complex parameters and
$\rho_{\vec{u}}:=\prod_{k=1}^{N} \otimes\rho_{u_k}\circ \Delta^{(N-1)}$, such that
\begin{eqnarray}
\rho_{\vec{u}}(\mathbf{u}_{0,n})=\sum_{k=1}^{N}\mathbf{1}\otimes\cdots\otimes\mathbf{1}\otimes\rho_{u_k}
(\mathbf{u}_{0,n}) \otimes\mathbf{1}\otimes\cdots\otimes\mathbf{1},
\end{eqnarray}
and
\begin{eqnarray}
\rho_{\vec{u}}(\mathbf{u}_{0,n})\cdot\prod_{k=1}^{N}\mathsf{Mac}_{\lambda^k}\{p^{(k)}\}=
\sum_{k=1}^{N}u_k^m\mathbf{C}_{\lambda^k}^{(m)}\cdot \prod_{k=1}^{N} \mathsf{Mac}_{\lambda^k}\{p^{(k)}\}.
\end{eqnarray}
In addition, due to the automorphism (\ref{automorphism}), there is another topological coproduct structure
$^{\sigma}\Delta=(\sigma\otimes\sigma)\circ\Delta\circ\sigma^{-1}$, then
\begin{eqnarray}
^{\sigma}\Delta(\mathbf{u}_{n,0})&=&\mathbf{u}_{n,0}\otimes\mathbf{1}+\mathbf{c}_{1,0}^n\otimes \mathbf{u}_{n,0},\nonumber\\
^{\sigma}\Delta(\mathbf{u}_{n,1})&=&\mathbf{u}_{n,1}\otimes\mathbf{1}+\sum_{k\ge0}\mathbf{c}_{1,0}^{n+k}
\mathbf{c}_{0,1}\theta_{-k,0}\otimes \mathbf{u}_{n+k,1},\\
^{\sigma}\Delta(\mathbf{u}_{n,-1})&=&\mathbf{u}_{n,-1}\otimes\mathbf{1}+\sum_{k\ge0}\mathbf{c}_{1,0}^{n-k}
\mathbf{c}_{0,1}^{-1}\theta_{k,0}\otimes \mathbf{u}_{n-k,-1},\nonumber
\end{eqnarray}
which means
\begin{eqnarray}
^{\sigma}\Delta\left(\sum_{n\in\mathbb{Z}}\mathbf{u}_{n,\pm1}z^n\right)=\left(\sum_{n\in\mathbb{Z}}
\mathbf{u}_{n,\pm1}z^n\right)\otimes\mathbf{1}+\left(\mathbf{c}_{0,1}^{\pm 1}\sum_{k\ge0}
\theta_{\mp k,0}z^{-k}\right)\otimes \left(\sum_{n\in\mathbb{Z}}\mathbf{c}_{1,0}^n\mathbf{u}_{n,\pm1}z^n\right).
\end{eqnarray}
	
By a direct calculation, we obtain
\begin{align}
X_{n}^+&:={\rho^\sigma_{\vec{u}}}(\mathbf{u}_{n,1}) =\frac{-1}{(1-q)(1-t^{-1})}
\sum_{k=1}^{N}u_k\oint\frac{\mathrm{d}z}{z^{n+1}}\Lambda_k^{+}(z),\nonumber\\
X_{n}^-&:={\rho^\sigma_{\vec{u}}}(\mathbf{u}_{n,-1}) =\frac{1}{(1-q^{-1})(1-t)}
\sum_{k=1}^{N}u_k^{-1}\oint\frac{\mathrm{d}z}{z^{n+1}}\Lambda_k^{-}(z), \label{X_0^+}
\end{align}
where
\begin{eqnarray}
^\sigma\Delta^{(1)}:&=&\mathbf{1},\nonumber\\
^\sigma\Delta^{(N)}:&=&(\overbrace{\mathbf{1}\otimes\cdots\otimes\mathbf{1}}^{N-2}\otimes{^\sigma\Delta})\circ\cdots\circ
(\mathbf{1}\otimes{^\sigma\Delta})\circ{^\sigma\Delta}, \qquad N\geq2,
\end{eqnarray}
$\rho^\sigma_{\vec{u}}:=\prod_{k=1}^{N} \otimes\rho_{u_k}\circ {^\sigma\Delta^{(N-1)}}$ and
$\Lambda_k^{\pm}(z)$ are given by
\begin{align}
\Lambda_k^{+}(z)&=\exp\left\{\sum_{n=1}^{\infty} \frac{1-t^{-n}}{n}p_n^{(k)}[(t/q)^{k-1\over2}z]^n\right\}\nonumber\\
&\cdot\exp\left\{-\sum_{n=1}^{\infty}(1-q^n)\left[\frac{\partial}{\partial p_n^{(k)}}+(1-(t/q)^n)
\sum_{l=1}^{k-1}(t/q)^{{l-k\over2}n} \frac{\partial}{\partial p_n^{(l)}}\right] [(t/q)^{k-1\over2}z]^{-n}\right\},\nonumber\\
\Lambda_k^{-}(z)&=\exp\left\{-\sum_{n=1}^{\infty} \frac{1-t^{-n}}{n}\left[(t/q)^n p_n^{(k)}-(1-(t/q)^n)
\sum_{l=1}^{k-1}(t/q)^{{l-k\over2}n}\frac{\partial}{\partial p_n^{(l)}}\right][(t/q)^{k-2\over2}z]^n\right\}\nonumber\\
&\cdot\exp\left\{-\sum_{n=1}^{\infty}(1-q^n)\frac{\partial}{\partial p_n^{(k)}} [(t/q)^{k-2\over2}z]^{-n}\right\}.
\end{align}

Note that
\begin{eqnarray}
\langle X_0^{+}f,g\rangle_{q,t} =\frac{-1}{(1-q)(1-t^{-1})}\langle f,X_0^{(1)}g\rangle_{q,t} \label{X_0^{(1)}},
\end{eqnarray}
for any symmetric functions $f$ and $g$, where the operator $X_0^{(1)}$ is given by \cite{Feigin09}
\begin{eqnarray}
X_{0}^{(1)}&=& \sum_{k=1}^{N}u_k\oint\frac{\mathrm{d}z}{z}\tilde{\Lambda}_k(z),
\end{eqnarray}
in which
\begin{eqnarray}
\tilde{\Lambda}_k(z)&=&\exp\left\{\sum_{n=1}^{\infty} \frac{1-t^{-n}}{n}[(t/q)^{k-1\over2}z]^n \left[p_n^{(k)}+(1-(t/q)^n)
\sum_{l=1}^{k-1}(t/q)^{{l-k\over2}n}p_n^{(l)}\right]\right\}\nonumber\\
&&\cdot\exp\left\{-\sum_{n=1}^{\infty}(1-q^n)\frac{\partial}{\partial p_n^{(k)}} [(t/q)^{k-1\over2}z]^{-n}\right\}.
\end{eqnarray}
$X_0^{(1)}$ is the standard operator used to define GMP \cite{Awata11,Fukuda20}.

Let us define the polynomials $M_{\vec{\lambda}}$ as follows:
\begin{subequations}
\begin{align}
&M_{\vec{\lambda}}\{\vec{u},\vec{p}\}=\prod_{1\le k<l\le N}G_{k,l}(\vec{\lambda};\vec{u})
\tilde{M}_{\vec{\lambda}}\{\vec{u},\vec{p}\},\\
&\tilde{M}_{\vec{\lambda}}\{\vec{u};\vec{p}\}=\prod_{k=1}^{N}\mathsf{Mac}_{\lambda^{k}}\{p^{(k)}\}
+\sum_{\vec{\mu}{<^R}\vec{\lambda}}\mathcal{V}^{\vec{\mu}}_{\vec{\lambda}}(\vec{u})\prod_{k=1}^{N}
\mathsf{Mac}_{\mu^k}\{p^{(k)}\}, \\
&X_0^+M_{\vec{\lambda}}\{\vec{u};\vec{p}\}=\sum_{k=1}^{N} u_k\mathbf{C}^{(1)}_{\lambda^{k}}
M_{\vec{\lambda}}\{\vec{u};\vec{p}\},\label{eigenMac}
\end{align}
\end{subequations}

where $\mathcal{V}^{\vec{\mu}}_{\vec{\lambda}}(\vec{u})$ is a complex matrix, and
\begin{align}
&G_{k,l}(\vec{\lambda};\vec{u})=G_{\lambda^{k},\lambda^{l}}(u_k/u_l),\qquad 1\le k,l\le N,\nonumber\\
&G_{\lambda,\mu}(x)= \prod_{(i,j)\in\lambda}(1-xq^{\lambda_i-j+1}t^{\mu'_j-i})\times
\prod_{(i,j)\in\mu}(1-xq^{-\mu_i+j}t^{-\lambda'_j+i-1}),\\
&\langle\mathsf{Mac}_{\lambda},\mathsf{Mac}_{\lambda}\rangle_{q,t}=(-1)^{|\lambda|}\prod_{(i,j)
\in\lambda}q^{j-1}t^{i} G_{\lambda,\lambda}(1),\nonumber
\end{align}
in particular,
\begin{eqnarray}
G_{\lambda,\emptyset}(x)&=&\prod_{(i,j)\in\lambda}(1-xq^jt^{-i}),\nonumber\\
G_{\emptyset,\lambda}(x)&=&\prod_{(i,j)\in\lambda}(1-xq^{1-j}t^{i-1}).
\end{eqnarray}
Note that our functions $G_{\lambda,\mu}(u)$ are identical to the functions $N_{\lambda,\mu}((q/t)u)$ in Ref.\cite{Awata11}.
	
The dual polynomial $M^*_{\vec{\lambda}}\{\vec{u} ,\vec{p}\}$ is defined by
\begin{eqnarray}
\langle M^*_{\vec{\mu}}\{\vec{u} ,\vec{p}\},M_{\vec{\lambda}}\{\vec{u},\vec{p}\}\rangle_{q,t}=\prod_{k,l=1}^{N}G_{k,l}
(\vec{\lambda};\vec{u})\delta_{\vec{\mu},\vec{\lambda}},\label{orthogonality}
\end{eqnarray}
which implies the Cauchy completeness identity
\begin{eqnarray}
\sum_{\vec{\lambda}} \frac{M_{\vec{\lambda}}\{\vec{u};\vec{p}\}M^*_{\vec{\lambda}}\{\vec{u};\vec{g}\}}{\prod_{k,l=1}^{N}G_{k,l}
(\vec{\lambda};\vec{u})}&=&\exp\left\{\sum_{k=1}^{N}\sum_{n\ge1}\frac{1-t^n}{1-q^n}\frac{p_n^{(k)}g_n^{(k)}}{n}\right\}.
\end{eqnarray}	
From (\ref{X_0^{(1)}}) and (\ref{orthogonality}), we see that $X^{(1)}_0$ can be diagonalized by $M^*_{\vec{\lambda}}\{\vec{u};\vec{p}\}$.
Thus, $M_{\vec{\lambda}}\{\vec{u};\vec{p}\}$ and $M^*_{\vec{\lambda}}\{\vec{u};\vec{p}\}$ correspond to the bra and ket
states of GMP, respectively \cite{Awata11}.
	
According to (\ref{theta1}), there are a series of operators commutating with $X_0^{+}$
\begin{eqnarray}
\mathcal{H}_n:=-\frac{1}{\kappa_1}\rho^{\sigma}_{\vec{u}}(\theta_{0,n})=
\begin{cases}
(t/q)^{-1/2}\mathrm{ad}_{X_{-1}^+}\mathrm{ad}_{X_0^+}^{n-2} X_1^+,& n\ge2,\\
X_0^+, & n=1, \\
X_0^-, & n=-1, \\
(t/q)^{1/2}\mathrm{ad}_{X_1^-}\mathrm{ad}_{X_0^-}^{n-2} X_{-1}^-,& n\le -2,
\end{cases}\label{qtHamilton}
\end{eqnarray}
and they can be regarded as the higher Hamiltonians for GMP.

Then, we have
\begin{eqnarray}
\mathcal{H}_nM_{\vec{\lambda}}\{\vec{u};\vec{p}\}&=&e^{(n)}_{\vec{\lambda}}M_{\vec{\lambda}}\{\vec{u};\vec{p}\},
\end{eqnarray}
where the eigenvalues are given by
\begin{eqnarray}
e_{\vec{\lambda}}^{(\pm n)}&=&
-\frac{1}{\kappa_1}\oint\frac{dz}{z^{1+n}}\prod_{k=1}^{N} P^{\pm}_{\lambda^k}(u_k z;q,t),
\end{eqnarray}
in which
\begin{eqnarray}
P_{\lambda}^+(z;q,t)&=&\exp\left\{-\sum_{m\ge1}\frac{\kappa_m}{m}\mathbf{C}_{\lambda^k}^{(m)} z^m\right\} \\ \nonumber
&=&\frac{1-zq^{-1}t}{1-z}\prod_{(i,j)\in\lambda} \frac{(1-q^jt^{-i}z)(1-q^{j-2}t^{1-i}z)(1-q^{j-1}t^{2-i}z)}
{(1-q^{j-1}t^{-i}z)(1-q^jt^{1-i}z)(1-q^{j-2}t^{2-i}z)}\\ \nonumber
&=&\frac{1-q^{\lambda_1-1}tz}{1-q^{\lambda_1}z}\prod_{i=1}^{\infty}\frac{(1-q^{\lambda_i}t^{-i}z)
(1-q^{\lambda_{i+1}}t^{-i+1}z)}{(1-q^{\lambda_{i+1}}t^{-i}z)(1-q^{\lambda_{i}-1}t^{-i+1}z)},\\
P_{\lambda}^-(z;q,t)&=&\frac{1}{P_{\lambda}^+(z;q^{-1},t^{-1})}.\nonumber
\end{eqnarray}
Here we have used a conjecture on the eigenfunctions
\begin{eqnarray}
\rho^{\sigma}_{\vec{u}}(\mathbf{u}_{0,\pm n}) M_{\vec{\lambda}}\{\vec{u};\vec{p}\}=
\pm\left(\sum_{k=1}^{N}u_k^{\pm}\mathbf{C}^{\pm n}_{\lambda^k}\right)
M_{\vec{\lambda}}\{\vec{u};\vec{p}\}, \label{Awataconjecture}
\end{eqnarray}
which is equivalent to the expression conjectured by Awata et al. in Ref.\cite{Awata16}.

\subsection{Limit to the $\beta$ deformed case}
In this subsection, we consider a certain limit to the $\beta$ deformed case.
Let us now consider the semi-classical limits: $t=q^{\beta}=e^{\beta\hbar}$, $u_k=q^{a_k}$ with $k=1,\cdots,N$
and taking $\hbar\rightarrow0$ with $\beta$ fixed. In this case, GMP degenerate to GJP.

Comparing (\ref{eigenGJP}) with (\ref{eigenMac}), we have
\begin{eqnarray}
\sum_{(i,j)\in\lambda}u_kq^{j-1}t^{1-i}&=& \sum_{(i,j)\in\lambda}e^{[a_k+(j-1)+(1-i)\beta]\hbar}\nonumber\\
&=&\sum_{n=0}^{\infty}\frac{\hbar^n}{n!}\sum_{(i,j)\in\lambda}[a_k+(j-1)+(1-i)\beta]^n.
\end{eqnarray}
Thus, we may rewrite the $\beta$-deformed operators (\ref{Woper0}) as
\begin{eqnarray}
W_{0,n+1}(\vec{a};\vec{p})=\sum_{k=1}^{N}W_{0,n+1}^{(k)}(\vec{a};\vec{p}):=n!\oint\frac{d\hbar}{\hbar^{n+1}}
\left(X_0^{+} +\frac{1}{(1-t^{-1})(1-q)} \sum_{k=1}^{N}q^{a_k}\right),\quad  n\ge0, \label{betaH}
\end{eqnarray}
which can be regarded as the higher Hamiltonians for GJP.

Let us list some operators
\begin{align}
W_{0,1}^{(k)}(\vec{a};\vec{p})&=\sum_{n\ge1}np_n^{(k)}\frac{\partial} {\partial p_n^{(k)}},\nonumber \\
W_{0,2}^{(k)}(\vec{a};\vec{p})&=\frac{1}{2}\sum_{n,m\ge1}\left[ \beta(n+m)p_n^{(k)}p_m^{(k)}
\frac{\partial}{\partial p_{n+m}^{(k)}} +nm p_{n+m}^{(k)}\frac{\partial^2}{\partial p_{n}^{(k)}\partial p_{m}^{(k)}}
+(1-\beta)(n-1)np_n^{(k)}\frac{\partial} {\partial p_n^{(k)}}\right] \nonumber \\
&+(1-\beta)\sum_{1\le l<k}\sum_{n\ge1}n^2 p_n^{(k)} \frac{\partial}{\partial p_n^{(l)}}
+ a_kW_{0,1}^{(k)}(\vec{a};\vec{p}),\nonumber\\
W_{0,3}^{(k)}(\vec{a};\vec{p})&=\sum_{n\ge1}\left[\frac{2n-1}{3}(1+\beta^2)-\frac{n-1}{2}
\beta\right]n(n-1)p_n^{(k)}\frac{\partial}{\partial p_n^{(k)}}\nonumber \\
&+\frac{1}{2}(1-\beta)\sum_{n,m\ge1}(n+m-1)\left[\beta(n+m)p_n^{(k)}p_m^{(k)}\frac{\partial}{\partial p_{n+m}^{(k)}}+nmp_{n+m}^{(k)}
\frac{\partial^2}{\partial p_n^{(k)}\partial p_m^{(k)}}\right]\nonumber \\
&+\frac{\beta^2}{3}\sum_{n,m,s\ge1}(n+m+s)p_n^{(k)}p_m^{(k)}p_s^{(k)}\frac{\partial}{\partial p_{n+m+s}^{(k)}}
+\frac{\beta}{2}\sum_{n+m=s+r}srp_{n}^{(k)}p_{m}^{(k)}\frac{\partial^2}
{\partial p_s^{(k)}\partial p_r^{(k)}}\nonumber\\
&+\frac{1}{3}\sum_{n,m,s\ge1}nmsp_{n+m+s}^{(k)}\frac{\partial^3}{\partial p_n^{(k)}\partial p_m^{(k)}
\partial p_s^{(k)}}+(1-\beta)^2\sum_{1\le l<k}\sum_{n\ge1}[n(k-l)-1]n^2p_n^{(k)}
\frac{\partial}{\partial p_n^{(l)}}\nonumber\\
&+(1-\beta)\sum_{1\le l<k}\sum_{n,m\ge1}(n+m)\left[\beta(n+m) p_n^{(k)}p_m^{(k)}
\frac{\partial}{\partial p_{n+m}^{(l)}}+nm p_{n+m}^{(k)}\frac{\partial^2}{\partial p_{n}^{(k)}
\partial p_{m}^{(l)}}\right]\nonumber\\
&+2a_kW_{0,2}^{(k)}(\vec{a};\vec{p})+a_k^2W_{0,1}^{(k)}(\vec{a};\vec{p}).
\end{align}
The first two operators coincide with the ones derived in Ref.\cite{Ohkuda}.

\subsection{Generalized $(q,t)$-deformed partition functions}

Let us construct the generalized $(q,t)$-deformed $W$-operators with the help of the recursive formulas (\ref{theta}) as follows:
\begin{subequations}\label{qtW}
\begin{align}
\exp\left\{-\sum_{l=1}^{\infty}\frac{\kappa_l}{l}\mathcal{W}_{\pm l,nl}^{\alpha}(\vec{u};\vec{m};\vec{p}) z^l\right\} &=
1-\kappa_1\sum_{s=1}^{\infty}[\rho^{\sigma}_{\vec{u}}(\mathbf{u}_{0,\pm1}),\operatorname{ad}^s_{\mathcal{W}_{\pm 1,n}^{\alpha}(\vec{u};\vec{m};\vec{p})}
\rho^{\sigma}_{\vec{u}}(\mathbf{u}_{0,\mp1})]z^s,\\
\mathcal{W}^{\alpha}_{\mp 1,n}(\vec{u};\vec{m};\vec{p})&=\prod_{i=1}^{n}\operatorname{ad}_{\mathcal{W}^{\alpha}(\vec{u};m_i;\vec{p})}
\rho^{\sigma}_{\vec{u}}(\mathbf{u}_{\pm1,0}),\qquad n\ge1,\\
\mathcal{W}_{0,r}(\vec{u};\vec{p})&=-\frac{r}{\kappa_r}\oint\frac{dz}{z^{r+1}}\operatorname{log}(1-\kappa_1\sum_{s\neq0}
\mathcal{H}_sz^s), \qquad r\neq0,\label{W0oper}
\end{align}
\end{subequations}
where $\alpha\in\{+,-\}$, $\mathcal{W}^{\pm}(\vec{u};m)=\rho^{\sigma}_{\vec{u}}(\mathbf{u}_{0,0}\mp m\mathbf{u}_{0,\pm1})$ and
\begin{eqnarray}\label{qtsigma}
\rho^{\sigma}_{\vec{u}}(\mathbf{u}_{l,0})=\begin{cases}
-\sum_{k=1}^{N}\frac{1}{1-q^l}(t/q)^{(k-2)l \over 2}p_l^{(k)},	& l>0,  \\
-\sum_{k=1}^{N}\frac{l}{1-t^{l}}(t/q)^{{(k-1)l \over 2}}\frac{\partial}{\partial p_{-l}^{(k)}}, & l<0.
\end{cases}
\end{eqnarray}
Note that a similar approach proposed in Ref.\cite{Fan23} to construct $(q,t)$-deformed $W$-operators is in terms of the
commutators of the $W$-operators and Macdonald difference operator. However its deficiency is clear.
Due to the commutators, it is hard to give exact expression of the $W$-operators.
As an improvement, our approach allows us to explicitly construct the $(q,t)$-deformed $W$-operators
based on the recursive formulas (\ref{theta}).
That is to say that the exact expression of the $W$-operators can be given recursively.
	
In order to give the explicit actions of these operators on GMP, we introduce the generalized $(q,t)$-deformed
cut-and-join rotation operators
\begin{subequations}\label{qtO}
\begin{eqnarray}
\hat{O}^+_{q,t}(\vec{u};x;\vec{p})&=&\exp\left\{-\sum_{n\ge1} \frac{x^n}{n}\left[\frac{1-(t/q)^n}
{\kappa_n}\sum_{k=1}^{N}u_k^n+\rho^{\sigma}_{\vec{u}}(\mathbf{u}_{0,n})\right]\right\}, \\
\hat{O}^-_{q,t}(\vec{u};x;\vec{p})&=&\exp\left\{\sum_{n\ge1} \frac{x^n}{n}\left[\frac{1-(t/q)^{-n}}{\kappa_{-n}}
\sum_{k=1}^{N}u_k^{-n}+\rho^{\sigma}_{\vec{u}}(\mathbf{u}_{0,-n})\right]\right\},
\end{eqnarray}
\end{subequations}
such that
\begin{eqnarray}
\hat{O}_{q,t}^{\alpha}(\vec{u};x;\vec{p}) M_{\vec{\lambda}}\{\vec{u};\vec{p}\}&=&
\mathcal{G}_{\vec{\lambda}}^{\alpha}(\vec{u};x)M_{\vec{\lambda}}\{\vec{u};\vec{p}\},
\end{eqnarray}
where
\begin{eqnarray}
\mathcal{G}_{\vec{\lambda}}^{+}(\vec{u};x)&=&\prod_{k=1}^{N} G_{\lambda^{k},\emptyset}((t/q)xu_k),\nonumber\\
\mathcal{G}_{\vec{\lambda}}^{-}(\vec{u};x)&=&\prod_{k=1}^{N} G_{\emptyset,\lambda^{k}}(x/u_k).
\end{eqnarray}

In terms of (\ref{qtsigma}) and (\ref{qtO}), the $W$-operators $\mathcal{W}^{\alpha}_{\pm l,nl}(\vec{u},\vec{m})$
can be represented as
\begin{eqnarray}
\mathcal{W}^{\alpha}_{-l,nl}(\vec{u};\vec{m};\vec{p})&=&\left(\prod_{i=1}^{n}\hat{O}^{\alpha}_{q,t}
(\vec{u};m_i;\vec{p})\right)\circ \rho^{\sigma}_{\vec{u}}(\mathbf{u}_{l,0})\circ
\left( \prod_{i=1}^{n}\hat{O}^{\alpha}_{q,t}(\vec{u};m_i;\vec{p})\right)^{-1}\nonumber\\
&=&-\sum_{k=1}^{N}\frac{(t/q)^{(k-2)l\over 2}}{1-q^l} \mathcal{W}^{\alpha,(k)}_{-l,nl}(\vec{u},\vec{m};\vec{p}),\nonumber\\
\mathcal{W}^{\alpha}_{l,nl}(\vec{u};\vec{m};\vec{p})&=&\left(\prod_{i=1}^{n}\hat{O}^{\alpha}_{q,t}(\vec{u};m_i;\vec{p})
\right)^{-1}\circ \rho^{\sigma}_{\vec{u}}(\mathbf{u}_{-l,0}) \circ \left(\prod_{i=1}^{n}\hat{O}^{\alpha}_{q,t}
(\vec{u};m_i;\vec{p})\right)\nonumber\\
&=&-\sum_{k=1}^{N}t^l\frac{(t/q)^{-{(k-1)l \over 2}}}{1-q^l}\mathcal{W}^{\alpha,(k)}_{l,nl}(\vec{u},\vec{m};\vec{p}). \label{qtwoper}
\end{eqnarray}

The actions of $W$-operators (\ref{qtwoper}) and $W_{0,\pm n}(\vec{u};\vec{p})$ (\ref{W0oper}) on GMP are
\begin{subequations}
\begin{eqnarray} \mathcal{W}^{\alpha}_{-l,nl}(\vec{u};\vec{m};\vec{p})M_{\vec{\lambda}}\{\vec{u};\vec{p}\}&=&
\sum_{|\vec{\mu}/\vec{\lambda}|=l}\mathcal{C}_{\vec{\lambda}}^{\vec{\mu}}\prod_{r=1}^{n}
\frac{\mathcal{G}_{\vec{\mu}}^{\alpha}(\vec{u};m_r)}{\mathcal{G}_{\vec{\lambda}}^{\alpha}(\vec{u};m_r)}
M_{\vec{\mu}}\{\vec{u};\vec{p}\},\\
\mathcal{W}^{\alpha}_{l,nl}(\vec{u};\vec{m};\vec{p})M_{\vec{\lambda}}\{\vec{u};\vec{p}\}&=&
\sum_{|\vec{\lambda}/\vec{\mu}|=l}\bar{\mathcal{C}}_{\vec{\lambda}}^{\vec{\mu}} \prod_{r=1}^{n}
\frac{\mathcal{G}_{\vec{\lambda}}^{\alpha}(\vec{u};m_r)}{\mathcal{G}_{\vec{\mu}}^{\alpha}(\vec{u};m_r)}
M_{\vec{\mu}}\{\vec{u};\vec{p}\},\\
\mathcal{W}_{0,\pm n}(\vec{u};\vec{p}) M_{\vec{\lambda}}\{\vec{u};\vec{p}\}&=&
\pm\sum_{k=1}^{N}\sum_{(i,j)\in\lambda^k}u_k^{\pm n}
\mathbf{C}_{\lambda^k}^{(\pm n)} M_{\vec{\lambda}}\{\vec{u};\vec{p}\},\quad n>0,
\end{eqnarray}
\end{subequations}
where the coefficient $\mathcal{C}_{\vec{\lambda}}^{\vec{\mu}}$ and $\bar{\mathcal{C}}_{\vec{\lambda}}^{\vec{\mu}}$
are given by
\begin{eqnarray}
\sum_{|\vec{\mu}/\vec{\lambda}|=l}\mathcal{C}_{\vec{\lambda}}^{\vec{\mu}} M_{\vec{\mu}}\{\vec{u};\vec{p}\}
&=&-\sum_{k=1}^{N}\frac{(t/q)^{(k-2)l\over 2}}{1-q^l}p_l^{(k)} M_{\vec{\lambda}}\{\vec{u};\vec{p}\} ,\nonumber\\
\sum_{|\vec{\lambda}/\vec{\mu}|=l}\bar{\mathcal{C}}_{\vec{\lambda}}^{\vec{\mu}} M_{\vec{\mu}}\{\vec{u};\vec{p}\}
&=&\sum_{k=1}^{N}l\frac{(t/q)^{-{(k-1)l \over 2}}}{1-t^{-l}}\frac{\partial}{\partial p_l^{(k)}} M_{\vec{\lambda}}\{\vec{u};\vec{p}\}.
\end{eqnarray}

With the construction of the $W$-operators we made above,  we may give the partition functions through $W$-representations
\begin{subequations}
\begin{eqnarray}
\mathcal{Z}_{-,n}(z;\vec{p};\vec{g})&=&\exp\left\{\sum_{l\ge1}\frac{z^l}{l}\frac{1-t^l}{1-q^l}\sum_{k=1}^{N}
\mathcal{W}^{+,(k)}_{-l,ln}(\vec{u};\vec{m};\vec{p})\mathcal{W}^{-,(k)}_{-l,ln}
(\vec{u};\vec{m}';\vec{g})\right\}\cdot1 \nonumber \\
&=&\sum_{\vec{\lambda}}z^{|\vec{\lambda}|}\prod_{k=1}^{N}\prod_{r=1}^{n}G_{\lambda^k,\emptyset}((t/q)m_ru_k)
G_{\emptyset,\lambda^k}(m'_r/u_k)\nonumber\\ &&\cdot\frac{M^*_{\vec{\lambda}}\{\vec{u};\vec{p}\}
M_{\vec{\lambda}}\{\vec{u};\vec{p}\}}{\prod_{k,l=1}^{N}G_{k,l}(\vec{\lambda})},\label{GMP} \\
\mathcal{Z}_{0,n}(z;\vec{p};\vec{g})&=&\exp\left\{\mathcal{W}_{0,\pm n}(\vec{u};\vec{p})\right\}
\cdot\exp\left\{\sum_{k=1}^{N}\sum_{l\ge1}\frac{z^l}{l}\frac{1-t^l}{1-q^l}p_l^{(k)}g_l^{(k)}\right\},\nonumber \\
&=&\sum_{\vec{\lambda}}z^{|\vec{\lambda}|}\exp\left(\pm\sum_{k=1}^{N}\sum_{(i,j)\in\lambda^k}
u_k^{\pm n}\mathbf{C}_{\lambda^k}^{(\pm n)}\right) \frac{M^*_{\vec{\lambda}}\{\vec{a};\vec{p}\}
M_{\vec{\lambda}}\{\vec{u};\vec{p}\}}{\prod_{k,l=1}^{N}G_{k,l}(\vec{\lambda})},\\
\mathcal{Z}_{+,n}(z,w;\vec{p};\vec{g})&=&\exp\left\{\sum_{l\ge1}\frac{z^l}{l}\frac{1-t^l}{1-q^l}
\sum_{k=1}^{N}\mathcal{W}^{+,(k)}_{l,nl}(\vec{u};\vec{m};\vec{p})\mathcal{W}^{-,(k)}_{l,nl}
(\vec{u};\vec{m}';\vec{g})\right\}\nonumber\\
&&\cdot\exp\left\{\sum_{k=1}^{N}\sum_{l\ge1}\frac{w^{-l}}{l}
\frac{1-t^l}{1-q^l}p_l^{(k)}g_l^{(k)}\right\}\nonumber \\
&=&\sum_{\vec{\mu}\subset\vec{\lambda}}\frac{z^{|\vec{\lambda}|}}{w^{|\vec{\mu}|}}\prod_{k=1}^{N}
\prod_{r=1}^{n} G_{\mu^k,\emptyset}((t/q)m_ru_k)G_{\emptyset,\mu^k}(m_r'/u_k)\frac{\prod_{k,l=1}^{N}G_{k,l}(\vec{\lambda})}
{\prod_{k,l=1}^{N}G_{k,l}(\vec{\mu})}\nonumber\\
&&\cdot M^*_{\vec{\lambda}/\vec{\mu}}\{\vec{u};\vec{p}\}
M_{\vec{\lambda}/\vec{\mu}}\{\vec{u};\vec{g}\},	
\end{eqnarray}
\end{subequations}
where the skew polynomials $M_{\vec{\lambda}/\vec{\mu}}$ are defined by
\begin{eqnarray}
\langle f,M_{\vec{\lambda}/\vec{\mu}} \rangle_{q,t}&=&\frac{1}{\langle M^*_{\vec{\mu}}, M_{\vec{\mu}}
\rangle_{\beta}}\langle fM^*_{\vec{\mu}}, M_{\vec{\lambda}}\rangle_{q,t},\nonumber\\
\langle f,M^*_{\vec{\lambda}/\vec{\mu}} \rangle_{q,t}&=&\frac{1}{\langle M^*_{\vec{\mu}}, M_{\vec{\mu}}
\rangle_{\beta}}\langle fM_{\vec{\mu}}, M^*_{\vec{\lambda}}\rangle_{q,t},
\end{eqnarray}
for any symmetric function $f$.

\subsection{$5d$ Nekrasov partition functions}
The $5d$ Nekrasov partition function for the $U(N)$ theory with $\mathcal{N}_f=2n$ fundamental
hypermultiplets is given by \cite{Awata5d,5d6d,Awata11}
\begin{eqnarray}
\mathcal{Z}_{inst}^{U(N),n}(z;\vec{u};\vec{m}^+,\vec{m}^-)=\sum_{\vec{\lambda}}z^{|\vec{\lambda}|}
\mathcal{Z}_{vec}(\vec{u},\vec{\lambda})\prod_{f=1}^n\mathcal{Z}_{fund}(\vec{u},\vec{\lambda};m_f^+)
\mathcal{Z}_{anti}(\vec{u},\vec{\lambda};m_f^-),\label{5dNek}
\end{eqnarray}
where
\begin{align}
\mathcal{Z}_{vec}(\vec{u},\vec{\lambda})&=\prod_{k,l=1}^{N}G^{-1}_{k,l}(\vec{\lambda}),\nonumber\\
\mathcal{Z}_{fund}(\vec{u},\vec{\lambda};m^+)&=\prod_{k=1}^{N}\prod_{(i,j)\in\lambda^{k}}(1-u_k^{-1}m^+q^{-j}t^i)\nonumber\\
&=\prod_{k=1}^{N}\left[G_{\lambda^{k},\emptyset}(u_k/m^+)\prod_{(i,j)\in\lambda^{k}}(-u_k^{-1}m^+q^{-j}t^i)\right],\nonumber\\
\mathcal{Z}_{anti}(\vec{u},\vec{\lambda};m^-)&=\prod_{k=1}^{N}\prod_{(i,j)\in\lambda^{k}}(1-u_km^-q^{j-1}t^{1-i})\nonumber \\
&=\prod_{k=1}^{N}\left[G_{\emptyset,\lambda^{k}}(u_k^{-1}/m^-)\prod_{(i,j)\in\lambda^{k}}(-u_km^-q^{j-1}t^{1-i})\right].
\end{align}

The $5d$ Nekrasov partition function can be given by vertex operators \cite{Awata11,Fukuda20}.
To do this, one first modifies the correspondence (\ref{inner}) as
\begin{eqnarray*}
\langle p_\lambda| p_\mu\rangle=\langle 0|\hat{a}_{-\lambda} \hat{a}_\mu|0\rangle=\langle p_\lambda,p_\mu\rangle_{q,t},
\end{eqnarray*}
where $[\hat{a}_n,\hat{a}_m]=n\frac{1-q^n}{1-t^n}\delta_{m+n,0}$.
	
There is the vertex operator \cite{Awata11}
\begin{eqnarray}
\phi_v^u(z)=\exp\left(-\sum_{n=1}^{\infty}\frac{1}{n}\frac{v^{n}-(t/q)^{n}u^{n}}{1-q^{n}}\hat{a}_{-n}z^{n}\right)
\exp\left(\sum_{n=1}^{\infty}\frac{1}{n}\frac{v^{-n}-u^{-n}}{1-q^{-n}}\hat{a}_{n}z^{-n}\right),
\end{eqnarray}
which satisfies a Pieri-type formula
\begin{eqnarray}
\langle \mathsf{Mac}_\lambda|\phi_v^u(z)| \mathsf{Mac}_\mu\rangle=G_{\lambda,\mu}(u/v)z^{|\lambda|-|\mu|}
(tv/q)^{|\lambda|}(-u/q)^{-|\mu|}t^{\sum_{(i,j)\in\lambda}(i-1)}q^{\sum_{(i,j)\in\mu}(j-1)}.
\end{eqnarray}
It is easy to obtain
\begin{eqnarray}
\mathcal{Z}^{U(1),1}_{inst}\left(\frac{qz_1}{tz_2};u;v,w^{-1}\right)&=&\langle 0|
\phi_u^w(z_2)\phi_v^u(z_1)|0 \rangle\nonumber\\
&=&\sum_{\lambda}\frac{\langle 0|\phi_u^w(z_2)| \mathsf{Mac}_\lambda\rangle\langle
\mathsf{Mac}_\lambda|\phi_v^u(z_1)| 0\rangle}
{\langle \mathsf{Mac}_\lambda| \mathsf{Mac}_\lambda\rangle}\nonumber\\
&=&\sum_{\lambda}\left(\frac{vz_1}{wz_2}\right)^{|\lambda|}
\frac{G_{\lambda,\emptyset}(u/v)G_{\emptyset,\lambda}(w/u)}{G_{\lambda,\lambda}(1)}.
\end{eqnarray}

More generally, there is the vertex operator (or Mukad\'e operator) $\Phi(z)$ in \cite{Fukuda20, Awata11} such that
\begin{eqnarray}
\langle M_{\vec{\lambda}}(\vec{u})|\Phi(z)| M_{\vec{\mu}}(\vec{v})\rangle
&=&(zt/q)^{|\vec{\lambda}|-|\vec{\mu}|} \left(\prod_{k=1}^N\frac{v_i}{u_i}\right)^{|\vec{\lambda}|}
\frac{\xi^{(+)}_{\vec{\lambda}}(\vec{u})}{\xi^{(+)}_{\vec{\mu}}(\vec{v})}\frac{\prod_{(i,j)
\in\mu^k}v_kq^{j-1}t^{1-i}}{\prod_{(i,j)\in\lambda^k}u_kq^{j-1}t^{1-i}}\nonumber\\
&&\cdot\prod_{i,j=1}^{N}G_{\lambda^i,\mu^j}(u_i/v_j),
\end{eqnarray}
where $\xi^{(+)}$ is defined as a normalized coefficient in \cite{Fukuda20}.
In terms of $\Phi(z)$, it is easy to give the $5d$ $U(N)$ Nekrasov partition function with $\mathcal{N}_{f}=2N$
fundamental matters
\begin{eqnarray}
\mathcal{Z}^{U(N),N}_{inst}\left(\frac{q^Nz_1}{t^Nz_2};\vec{u};\vec{v},\vec{w}^{-1}\right)&=&
\langle0|\Phi^{\vec{w}}_{\vec{u}}(z_{2})\Phi^{\vec{u}}_{\vec{v}}(z_{1})|0\rangle\nonumber\\
&=&\sum_{\vec{\lambda}} \frac{\langle 0|\Phi_{\vec{u}}^{\vec{w}} (z_{2})|M_{\vec{\lambda}}\rangle\langle
M_{\vec{\lambda}}|\Phi_{\vec{v}}^{\vec{u}}(z_{1})|0\rangle}{\langle M_{\vec{\lambda}}|M_{\vec{\lambda}}\rangle}\nonumber\\
&=&\sum_{\vec{\lambda}}\left(\frac{z_{1}}{z_{2}}\prod_{k=1}^{N}\frac{v_i}{w_i}\right)^{|\vec{\lambda}|}
\prod_{i,j=1}^{N}\frac{G_{\lambda^i,\emptyset}(u_{i}/v_{j})G_{\emptyset,\lambda^i}(w_{j}/u_{i})}
{G_{\lambda^i,\lambda^j}(u_{i}/u_{j})},\label{vertex}
\end{eqnarray}
where $\vec{w}^{-1}=(w_1^{-1},w_2^{-1},\cdots)$.

Going back to our partition functions (\ref{GMP}), we observe that by removing the generalized Jack polynomials
from $\mathcal{Z}_{-,n}$ (\ref{GMP}) and taking  $m_i=(q/t)(m^+_i)^{-1}$, $m'_i=(m^-_i)^{-1}$ with $i=1,2,\cdots,n$,
the remains match with the $5d$ Nekrasov partition functions  with $\mathcal{N}_{f}$ fundamental matters (\ref{5dNek}).
With this observation, we may establish the connection between these two partition functions	
\begin{eqnarray}
\mathcal{Z}_{inst}^{U(N),n}(z;\vec{u};\vec{m}^+,\vec{m}^-)&=&\left\langle\sum_{\vec{\lambda}}
\frac{M^*_{\vec{\lambda}}\{\vec{u};\vec{g}\}M_{\vec{\lambda}}\{\vec{u};\vec{p}\}}{\prod_{k,l=1}^{N}G_{k,l}^2
(\vec{\lambda})},\mathcal{Z}_{-,n}\{(q/t)^nz;\vec{p};\vec{g}\}\right\rangle_{q,t}. \label{vertexwope}
\end{eqnarray}

Then we make the explicit connection between the $W$-operators (\ref{qtwoper}) and vertex operator $\Phi(z)$
from (\ref{vertex}) and (\ref{vertexwope})
\begin{eqnarray}
&&\left\langle\sum_{\vec{\lambda}}
\frac{M^*_{\vec{\lambda}}\{\vec{u};\vec{g}\}M_{\vec{\lambda}}\{\vec{u};\vec{p}\}}{\prod_{k,l=1}^{N}G_{k,l}^2
(\vec{\lambda})},\exp\left\{\sum_{l\ge1}\mathcal{A}_l\sum_{k=1}^{N}
\mathcal{W}^{+,(k)}_{-l,lN}(\vec{u};\vec{m};\vec{p})\mathcal{W}^{-,(k)}_{-l,lN}(\vec{u};\vec{m}';\vec{g})
\right\}\cdot1\right\rangle_{q,t}\nonumber \\
&=&\langle0|\Phi^{\vec{w}}_{\vec{u}}(z_{2})\Phi^{\vec{u}}_{\vec{v}}(z_{1})|0\rangle,
\end{eqnarray}	
where $\mathcal{A}_l=\frac{1}{l}\left(\frac{q^Nz_1}{t^Nz_2}\right)^l\frac{1-t^l}{1-q^l}$
and  $m_i=(q/t)(v_i)^{-1}$, $m'_i=w_i$ for $i=1,2,\cdots,N$.

\section{Conclusion}
We have attempted to construct the generalized $\beta$ and $(q,t)$-deformed partition functions through $W$-representations,
where the expansions are respectively with respect to the GJP and GMP labeled by $N$-tuple of Young diagrams.
To achieve the desired results, the key point is the construction of $W$-operators.
Based on the $N$-Fock representations of the $\mathbf{SH^c}$ and elliptic Hall algebras, we constructed generalized
$\beta$ and $(q,t)$-deformed $W$-operators, respectively.
In order to present the actions of the $W$-operators on the GJP and GMP more explicitly,
we have introduced the generalized cut-and-join rotation operators. In addition,
we presented the higher Hamiltonians (\ref{qtHamilton}) for GMP. Furthermore, the Hamiltonians for GJP (\ref{betaH})
can be obtained by the semi-classical limits of the operator $X^{+}_0$ (\ref{X_0^+}). Unfortunately, we failed to prove
the eigenvalue conjecture (\ref{Awataconjecture}) which is equivalent to the expression conjectured by Awata et al.
in Ref.\cite{Awata16}. We leave this as a future work.

As the particular case, based on the Fock representation (symmetric functions representation) of the algebra $\mathbf{SH^c}$,
we also constructed the $\beta$-deformed $W$-operators and presented the $\beta$-deformed partition functions
through $W$-representations. It was noted that the $\beta$-deformed $W$-operators are closely related to one-dimensional
many-body systems, such as the classical and $U(L|M)$-deformed rCS models. In addition, we found that the $\beta$-deformed ABJ-like
model (\ref{ABJ}) is the integral realization of $Z_{-2}(L_{eff};\bar{p})$ in (\ref{Neg1}).

By means of the desired deformed $W$-operators, we have achieved the generalized $\beta$ and $(q,t)$-deformed partition functions
through $W$-representations. The most interesting result is that there are the much profound interrelations between our deformed
partition functions and the $4d$ and $5d$ Nekrasov partition functions. We also noticed that the free energy of the Nekrasov
partition function $Z_{inst}^{U(1),1}(s;0;u;v)|_{\beta=1}$ (\ref{dessin}) can be given by the specific part of (connected)
Belyi fat graphs. It is well-known that the corresponding Nekrasov partition functions can be given by vertex operators.
Thus the deep connection between our $\beta$ and $(q,t)$-deformed $W$-operators and vertex operators was revealed in this paper.
Owing to this remarkable connection, more applications of these deformed $W$-operators would merit further investigations.
For further research, it would also be interesting to search for the integral representations for the generalized $\beta$
and $(q,t)$-deformed partition functions.

\section* {Acknowledgements}

This work is supported by the National Natural Science Foundation of China (Nos. 12375004 and 12205368 )
and the Fundamental Research Funds for the Central Universities, China (No. 2024ZKPYLX01).


\begin{thebibliography}{}

\bibitem{Nekrasov}
N.A. Nekrasov,
Seiberg-Witten prepotential from instanton counting,
Adv. Theor. Math. Phys. \textbf{7} (2004) 831 [hep-th/0206161].

\bibitem{Nekrasov2}
N. Nekrasov and A. Okounkov,
Seiberg-Witten theory and random partitions,
Prog. Math. \textbf{244} (2006) 525 [hep-th/0306238]

\bibitem{AGT}
L.F. Alday, D. Gaiotto and Y. Tachikawa,
Liouville correlation functions from four-dimensional gauge theories,
Lett. Math. Phys. $\mathbf{91}$ (2010) 167 [arXiv:0906.3219].

\bibitem{Awata5d}
H. Awata and Y. Yamada,
Five-dimensional AGT relation and the deformed $\beta$-ensemble,
Prog. Theor. Phys. \textbf{124} (2010) 227 [arXiv:1004.5122].

\bibitem{5d6d}
M.C. Tan,
An M-theoretic derivation of a $5d$ and $6d$ AGT correspondence, and relativistic and elliptized integrable systems,
JHEP $\mathbf{12}$ (2013) 031 [arXiv: 1309.4775].

\bibitem{SUN}
N. Wyllard,
Instanton partition functions in $N=2$ $SU(N)$ gauge theories with a general surface operator and their $W$-algebra duals,
JHEP $\mathbf{02}$ (2011) 114 [arXiv:1012.1355].

\bibitem{AFLT}
V.A. Alba, V.A. Fateev, A.V. Litvinov and G.M. Tarnopolskiy,
On combinatorial expansion of the conformal blocks arising from AGT conjecture,
Lett. Math. Phys. $\mathbf{98}$ (2011) 33 [arXiv:1012.1312].

\bibitem{Fateev12}
V.A. Fateev and A. V. Litvinov,
Integrable structure, $W$-symmetry and AGT relation,
JHEP $\mathbf{01}$ (2012) 051 [arXiv:1109.4042].

\bibitem{Zhang13}
S. Kanno, Y. Matsuo and H. Zhang,
Extended conformal symmetry and recursion formulae for Nekrasov partition function,
JHEP $\mathbf{08}$ (2013) 028 [arXiv:1306.1523].

\bibitem{Zhang14}
Y. Matsuo, C. Rim and H. Zhang,
Construction of Gaiotto states with fundamental multiplets through degenerate DAHA,
JHEP $\mathbf{09}$ (2014) 028 [arXiv: 1405.3141].

\bibitem{Bourgine}
J.E. Bourgine, M. Fukuda, Y. Matsuo, H. Zhang and R. D. Zhu,
Coherent states in quantum $W_{1+\infty}$ algebra and $qq$-character for 5$d$ super Yang-Mills,
Prog. Theor. Exp. Phys. $\mathbf{2016}$ (2016) 123B05 [arXiv:1606.08020].

\bibitem{Awata11}
H. Awata,  B. Feigin, A. Hoshino, M. Kanai, J. Shiraishi and S. Yanagida,
Notes on Ding-Iohara algebra and AGT conjecture,
RIMS kokyuroku $\mathbf{1765}$ (2011) 12 [arXiv:1106.4088].

\bibitem{Fukuda20}
M. Fukuda, Y. Ohkubo and J. Shiraishi,
Generalized Macdonald functions on Fock tensor spaces and duality formula for changing preferred direction,
Commun. Math. Phys. $\mathbf{380}$ (2020) 1 [arXiv:1903.05905].

\bibitem{Smir}
A. Smirnov,
Polynomials associated with fixed points on the instanton moduli space,
arXiv:1404.5304.

\bibitem{Tsymbaliuk17}
A. Tsymbaliuk,
The affine Yangian of $\mathfrak{gl}_1$ revisited,
Adv. Math. \textbf{304} (2017) 583 [arXiv:1404.5240].

\bibitem{Schiffmann13}
O. Schiffmann and E. Vasserot,
Cherednik algebras, $W$-algebras and the equivariant cohomology of the moduli space of instantons on $\mathbb{A}^2$,
Publ. Math. Inst. Hautes \'Etudes Sci. $\mathbf{118}$ (2013) 213 [arXiv:1202.2756].

\bibitem{Feigin12}
B. Feigin, E. Feigin, M. Jimbo, T. Miwa and E. Mukhin,
Quantum toroida $\mathfrak{gl}_{1}$-algebra: Plane partitions,
Kyoto J. Math. $\mathbf{52}$ (2012) 621 [arXiv:1110.5310].

\bibitem{DF}
V. Dotsenko and V. Fateev,
Conformal algebra and multipoint correlation functions in two-dimensional statistical models,
Nucl. Phys. B \textbf{240} (1984) 312.

\bibitem{Mir11}
A. Mironov, A. Morozov and Sh. Shakirov, A direct proof of AGT conjecture at $\beta=1$,
JHEP $\mathbf{02}$ (2011) 067 [arXiv:1012.3137].

\bibitem{Morozov14}
A. Morozov and A. Smirnov,
Towards the proof of AGT relations with the help of the generalized Jack polynomials,
Lett. Math. Phys. $\mathbf{104}$ (2014) 585 [arXiv: 1307.2576].

\bibitem{SMironov}
S. Mironov, A. Morozov and Y. Zenkevich,
Generalized Jack polynomials and the AGT relations for the $SU(3)$ group,
JETP Lett. $\mathbf{99}$ (2014) 115 [arXiv:1312.5732].

\bibitem{Zenkevich15}
Y. Zenkevich,
Generalized Macdonald polynomials, spectral duality for conformal blocks
and AGT correspondence in five dimensions,
JHEP $\mathbf{05}$ (2015) 131 [arXiv:1412.8592].

\bibitem{Zenkevich16}
A. Morozov and Y. Zenkevich,
Decomposing Nekrasov decomposition,
JHEP $\mathbf{02}$ (2016) 098 [arXiv:1510.01896].

\bibitem{Zhang11}
H. Zhang and Y. Matsuo,
Selberg integral and $SU(N)$ AGT conjecture,
JHEP $\mathbf{12}$ (2011) 106 [arXiv:1110.5255].

\bibitem{Itoyama}
H. Itoyama, K. Maruyoshi and T. Oota,
The quiver matrix model and 2$d$-4$d$ conformal connection,
Prog. Theor. Phys. \textbf{123} (2010) 957 [arXiv:0911.4244].

\bibitem{MM12}
A. Mironov, A. Morozov and S. Shakirov,
Towards a proof of AGT conjecture by methods of matrix models,
Int. J. Mod. Phys. A \textbf{27} (2012) 1230001 [arXiv:1011.5629].

\bibitem{Nishinaka}
T. Nishinaka and C. Rim,
$\beta$-deformed matrix model and Nekrasov partition function,
JHEP \textbf{02} (2012) 114 [arXiv:1112.3545].

\bibitem{Tai}
T. Tai,
Instanton counting and matrix model,
Prog. Theor. Phys. \textbf{119} (2008) 165 [arXiv:0709.0432].

\bibitem{Sulkowski}
P. Sulkowski,
Matrix models for $\beta$-ensembles from Nekrasov partition functions,
JHEP \textbf{04} (2010) 063 [arXiv:0912.5476]

\bibitem{Bonelli}
G. Bonelli, K. Maruyoshi, A. Tanzini and F. Yagi,
Generalized matrix models and AGT correspondence at all genera,
JHEP \textbf{07} (2011) 055.

\bibitem{MirPRD}
A. Mironov and A. Morozov,
Superintegrability as the hidden origin of the Nekrasov calculus,
Phys. Rev. D $\mathbf{106}$ (2022) 126004 [arXiv:2207.08242].


\bibitem{Morozov09}
A. Morozov and Sh. Shakirov,
Generation of matrix models by $\hat{W}$-operators,
JHEP $\mathbf{04}$ (2009) 064 [arXiv: 0902.2627].

\bibitem{Kon-Wit}
A. Alexandrov,
Cut-and-Join operator representation for Kontsewich-Witten tau-function,
Mod. Phys. Lett. A $\mathbf{26}$ (2011) 2193 [arXiv:1009.4887].

\bibitem{BGW}
A. Alexandrov,
Cut-and-join description of generalized Brezin-Gross-Witten model,
Adv. Theor. Math. Phys. $\mathbf{22}$ (2018) 1347 [arXiv:1608.01627].

\bibitem{Cassia}
L. Cassia, R. Lodin and M. Zabzine,
On matrix models and their $q$-deformations,
JHEP $\mathbf{10}$ (2020) 126 [arXiv:2007.10354].

\bibitem{MirGKM}
A. Mironov, V. Mishnyakov and A. Morozov,
Non-Abelian $W$-representation for GKM,
Phys. Lett. B $\mathbf{823}$ (2021) 136721 [arXiv:2107.02210].

\bibitem{Goulden01}
I. Goulden, D. Jackson and R. Vakil,
The Gromov-Witten potential of a point, Hurwitz numbers, and Hodge integral,
Proc. London Math. Soc. \textbf{83} (2001) 563 [math/9910004].

\bibitem{Al1405}
A. Alexandrov, A. Mironov, A. Morozov and S. Natanzon, On KP-integrable Hurwitz functions,
JHEP $\mathbf{11}$ (2014)  080 [arXiv:1405.1395].

\bibitem{Rui22}
R. Wang, F. Liu, C.H. Zhang and W.Z. Zhao,
Superintegrability for ($\beta$-deformed) partition function hierarchies with $W$-representations,
Eur. Phys. J. C $\mathbf{82}$ (2022) 902 [arXiv:2206.13038].

\bibitem{Alexandrov23}
A. Alexandrov,
On $W$-operators and superintegrability for dessins d'enfant,
Eur. Phys. J. C  $\mathbf{83}$ (2023) 147 [arXiv:2212.10952].

\bibitem{MironovSkew}
A. Mironov, V. Mishnyakov, A. Morozov, A. Popolitov and W.Z. Zhao,
On KP-integrable skew Hurwitz $\tau$-functions and their $\beta$-deformations,
Phys. Lett. B \textbf{839} (2023) 137805 [arXiv:2301.11877].

\bibitem{MironovIn}
A. Mironov, V. Mishnyakov, A. Morozov, A. Popolitov, R. Wang and W.Z. Zhao,
Interpolating matrix models for WLZZ series,
Eur. Phys. J. C $\mathbf{83}$ (2023) 377 [arXiv:2301.04107].

\bibitem{Oreshina1}
A. Mironov, A. Oreshina and A. Popolitov,
Two $\beta$-ensemble realization of $\beta$-deformed WLZZ models,
arXiv:2403.05965.

\bibitem{Oreshina2}
A. Mironov, A. Oreshina and A. Popolitov,
$\beta$-WLZZ models from $\beta$-ensemble integrals direactly,
arXiv:2404.18843.

\bibitem{MironovCom}
A. Mironov, V. Mishnyakov, A. Morozov and A. Popolitov,
Commutative families in $W_{\infty}$, integrable many-body systems and hypergeometric $\tau$-functions,
JHEP \textbf{09} (2023) 065 [arXiv:2306.06623].

\bibitem{szabo1}
R.J. Szabo and M. Tierz,
$q$-deformations of two-dimensional Yang-Mills theory: Classification, categorification and refinement,
Nucl. Phys. B \textbf{876} (2013) 234 [arXiv:1305.1580].

\bibitem{szabo2}
Z. K\"ok\'enyesi, A. Sinkovics and R.J. Szabo,
Refined Chern-Simons theory and ($q,t$)-deformed Yang-Mills theory: Semi-classical expansion and planar limit,
JHEP \textbf{10} (2013) 067 [arXiv:1306.1707].

\bibitem{AwataNet}
H. Awata, H. Kanno, T. Matsumoto, A. Mironov, Alexei Morozov, Andrey Morozov, Y. Ohkubo and Y. Zenkevich,
Explicit examples of DIM constraints for network matrix models,
JHEP $\mathbf{07}$ (2016) 103 [arXiv:1604.08366].

\bibitem{ZenkevichNet}
A. Mironov,  A. Morozov and Y. Zenkevich,
Ding-Iohara-Miki symmetry of network matrix models,
Phys. Lett. B $\mathbf{762}$ (2016) 196 [arXiv:1603.05467].

\bibitem{Mor18}
A. Morozov, A. Popolitov and Sh. Shakirov,
On ($q,t$)-deformation of Gaussian matrix model,
Phys. Lett. B \textbf{784} (2018) 342 [arXiv:1803.11401].

\bibitem{Mor19}
A. Morozov,
On $W$-representations of $\beta$- and $q,t$-deformed matrix models,
Phys. Lett. B \textbf{792} (2019) 205 [arXiv:1901.02811]

\bibitem{Morell}
A. Morozov and A. Mironov,
Elliptic $q,t$ matrix models,
Phys. Lett. B \textbf{816} (2021) 136196 [arXiv:2011.01762].

\bibitem{Morell2}
A. Morozov and A. Mironov,
Towards elliptic deformation of $q,t$ matrix models,
Phys. Lett. B \textbf{816} (2021) 136221 [arXiv:2011.02855].

\bibitem{Fan23}
F. Liu, A. Mironov, V. Mishnyakov, A. Morozov, A. Popolitov, R. Wang and W.Z. Zhao,
$(q,t)$-deformed (skew) Hurwitz $\tau$-functions,
Nucl. Phys. B $\mathbf{993}$ (2023) 116283 [arXiv:2303.00552].

\bibitem{Bou23}
J.E. Bourgine and A. Garbali,
A $(q,t)$-deformation of the 2$d$ Toda integrable hierarchy,
arXiv:2308.16583.

\bibitem{Feigin09}
B. Feigin, K. Hashizume, A. Hoshino, J. Shiraishi and S. Yanagida,
A commutative algebra on degenerate $\mathbb{CP}^{1}$ and Macdonald polynomials,
J. Math. Phys. $\mathbf{50}$ (2009) 095215 [arXiv:0904.2291].

\bibitem{D-I}
J. Ding and K. Iohara,
Generalization and deformation of Drinfeld quantum affine algebras,
Lett. Math. Phys. $\mathbf{41}$ (1997) 181 [q-alg/9608002].

\bibitem{Miki07}
K. Miki,
A $(q,\gamma)$ analog of the $W_{1+\infty}$ algebra,
J. Math. Phys. $\mathbf{48}$ (2007) 3520.

\bibitem{Feigin17}
B. Feigin, M. Jimbo, T. Miwa and E. Mukhin,
Finite type modules and Bethe ansatz for quantum toroidal $\mathfrak{gl}_1$,
Commun. Math. Phys. $\mathbf{356}$ (2017) 285 [arXiv:1603.02765].

\bibitem{Bourgine17}
J.E. Bourgine, M. Fukuda, K. Harada, Y. Matsuo and R.D. Zhu,
$(p,q)$-webs of DIM representations, $5d$ $N=1$ instanton partition functions and $qq$-characters,
JHEP \textbf{11} (2017) 034 [arxiv:1703.10759].

\bibitem{Awata17}
H. Awata, H. Fujino, and Y. Ohkubo,
Crystallization of deformed Virasoro algebra, Ding-Iohara-Miki algebra, and $5D$ AGT correspondence,
J. Math. Phys. \textbf{58} (2017) 071704.

\bibitem{SchiffmannHall}
I. Burban and O. Schiffmann,
On the Hall algebra of an elliptic curve, I,
Duke Math. J. $\mathbf{161}$ (2012) 1171 [math/0505148].

\bibitem{SchiffmannKA}
O. Schiffmann and E. Vasserot,
The elliptic Hall algebra and the K-theory of the Hilbert scheme of $\mathbb{A}^2$,
Duke. Math. J. $\mathbf{162}$ (2013) 279 [arXiv:0905.2555].

\bibitem{Prochazka}
T. Proch\'azka,
$W$-symmetry, topological vertex and affine Yangian,
JHEP $\mathbf{10}$ (2016) 077 [arXiv:1512.07178].

\bibitem{Nakajima}
H. Nakajima,
More lectures on Hilbert schemes of points on surfaces,
arXiv:1401.6782.

\bibitem{Lehn}
M. Lehn,
Chern classes of tautological sheaves on Hilbert schemes of points on surfaces,
Invent. Math. \textbf{136} (1999) 157 [math/9803091].

\bibitem{Sergeev05}
A.N. Sergeev and A.P. Veselov,
Generalised discriminants, deformed Calogero-Moser -Sutherland operators and super-Jack polynomials,
Adv. Math. $\mathbf{192}$ (2005) 341 [math-ph/0307036].

\bibitem{Macdonald}
I.G. Macdonald,
Symmetric functions and Hall polynomials,
Oxford Science Publications, Oxdford, U.K., (1995).

\bibitem{Fuhao}
F.H. Zhang, F. Liu, Y. Li and C.H. Zhang,
On higher deformed Calogero-Sutherland Hamiltonians, submitted.

\bibitem{Cassia21}
L. Cassia, R. Lodin and M. Zabzine,
Virasoro constraints revisited,
Commun. Math. Phys. \textbf{387} (2021) 1729 [arXiv:2102.05682].

\bibitem{CO}
E. Carlsson and A. Okounkov,
Exts and vertex operators,
Duke Math. J. $\mathbf{161}$ (2012) 1797 [arXiv:0801.2565].

\bibitem{ZhouJ}
J. Zhou,
Grothendieck's dessins d'enfants in a web of dualities,
arXiv:1905.10773.

\bibitem{Zenkevich23}
Y. Zenkevich,
On pentagon identity in Ding-Iohara-Miki algebra,
JHEP $\mathbf{03}$ (2023) 193 [arXiv:2112.14687].

\bibitem{Awata16}
H. Awata, H. Kanno, A. Mironov, Alexei Morozov, Andrey Morozov, Y. Ohkubo and Y. Zenkevich,
Toric Calabi-Yau threefolds as quantum integrable systems. $R$-matrix and $RTT$ relations,
JHEP $\mathbf{10}$ (2016) 047 [arXiv:1608.05351].	

\bibitem{Ohkuda}
Y. Ohkubo,
Generalized Jack and Macdonald polynomials arising from AGT conjecture,
J. Phys. Conf. Ser. $\mathbf{804}$ (2017) 012036 [arXiv: 1404.5401].
\end{thebibliography}
\end{document}